\documentclass[%
 aip,
 amsmath,amssymb,
 reprint,%
]{revtex4-2}

\usepackage{graphicx}
\usepackage{epstopdf}
\usepackage{dcolumn}
\usepackage{bm}

\usepackage[utf8]{inputenc}
\usepackage[T1]{fontenc}
\usepackage{mathptmx}
\usepackage{etoolbox}
\usepackage{color}
\usepackage{url}
\usepackage[colorlinks=true, linkcolor=blue, citecolor=blue, urlcolor=blue]{hyperref}

\makeatletter
\def\@email#1#2{%
 \endgroup
 \patchcmd{\titleblock@produce}
  {\frontmatter@RRAPformat}
  {\frontmatter@RRAPformat{\produce@RRAP{*#1\href{mailto:#2}{#2}}}\frontmatter@RRAPformat}
  {}{}
}%
\makeatother
\begin{document}

\preprint{AIP/123-QED}

\title{Quantum Biology, Quantum Simulation and Quantum Coherent Devices}

\author{Rong-Hang Chen}
\affiliation{Beijing Computational Science Research Center, Beijing 100193, China}
\affiliation{School of Physics and Astronomy, Applied Optics Beijing Area Major Laboratory, Beijing Normal University, Beijing 100875, China}
\affiliation{Key Laboratory of Multiscale Spin Physics, Ministry of Education, Beijing Normal University, Beijing 100875, China}

\author{Jing Dong}
\affiliation{School of Physics and Astronomy, Applied Optics Beijing Area Major Laboratory, Beijing Normal University, Beijing 100875, China}
\affiliation{ Institute of Physics, Chinese Academy of Sciences, Beijing 100190, China}
\affiliation{ University of Chinese Academy of Sciences, Beijing 100049, China}
\affiliation{Key Laboratory of Multiscale Spin Physics, Ministry of Education, Beijing Normal University, Beijing 100875, China}

\author{Wen Yang}
\affiliation{Beijing Computational Science Research Center, Beijing 100193, China}

\author{Qing Ai}
\thanks{Electronic mail: aiqing@bnu.edu.cn}
\affiliation{School of Physics and Astronomy, Applied Optics Beijing Area Major Laboratory, Beijing Normal University, Beijing 100875, China}
\affiliation{Key Laboratory of Multiscale Spin Physics, Ministry of Education, Beijing Normal University, Beijing 100875, China}

\author{Gui-Lu Long}
\thanks{Electronic mail: gllong@tsinghua.edu.cn}
\affiliation{Beijing Key Laboratory of Fault-Tolerant Quantum Computing, Beijing Academy of Quantum Information Sciences, Beijing 100193, China}
\affiliation{State Key Laboratory of Low-Dimensional Quantum Physics and Department of Physics, Tsinghua University, Beijing 100084, China}
\affiliation{Frontier Science Center for Quantum Information, Beijing 100084, China}
\affiliation{Beijing National Research Center for Information Science and Technology, Beijing 100084, China}

\date{\today}
	
\begin{abstract}
Many living organisms can exploit quantum mechanical effects to gain distinct biological advantages. In plants, photosynthesis uses quantum coherence to achieve near 100\% efficiency in energy transfer. With advances in experimental techniques, two-dimensional electronic spectroscopy can reveal dynamic processes such as coherence and coupling within a system, and it plays an important role in studying energy transfer in photosynthesis. On the theory side, methods such as the generalized Bloch–Redfield theory and the hierarchical equations of motion are used to model photosynthetic systems. Quantum simulation, as a high-efficiency and low-complexity approach, has also made progress across various platforms in the study of photosynthesis. In recent years, a series of studies has introduced quantum coherence into artificial systems to enhance energy transfer efficiency, laying the groundwork for the design of coherent devices with efficient energy transport.
Birds can use the weak geomagnetic field and spin-dependent chemical reactions to detect direction. Theoretical frameworks for animal navigation include magnetite-based mechanisms, magnetoreceptor genes, and the radical-pair mechanism. Quantum simulations of navigation have also advanced on multiple platforms. Inspired by animal navigation, diverse quantum effects have been applied to improve sensing and to support navigation tasks.
This paper presents a comprehensive review of progress on quantum coherence in photosynthesis and avian navigation, along with related theoretical methods, quantum simulation approaches, and research on quantum coherent devices.

\end{abstract}
\maketitle
	
\section{Introduction}	
	



Quantum coherence effects are widespread in living organisms, such as in the energy transfer process of photosynthesis and in the magnetic sensing used for animal navigation \cite{CaoJianshu2020SA, LambertNeill2013NP}. Photosynthesis is a complex biochemical reaction process. It is well known that the process of photosynthesis mainly includes four aspects, namely the primary reaction, electron transfer, photophosphorylation, and carbon dioxide fixation \cite{Fleming1994PT}. Due to nearly 100\% energy transfer efficiency, exciton energy transfer (EET) is one of the current research hotspots. However, the processes of energy and electron transfer are ultrafast, and both occur on the order of 1 ps. The excitation energy is transferred to the reaction center on a timescale of 30 ps, and the subsequent electron transfer is completed in about 3 ps. Since the energy transfer process in photosynthesis can be regarded as an open quantum system \cite{Melih2011CPC, Forster1948AP,  NovoderezhkinVladimirI2007BJ, NovoderezhkinVladimirI2010PCCP, SchroderMarkus2006JCP, JeskeJan2015JCP, LiuHao2014JCP}, there are theoretical methods such as F{\"o}rster theory \cite{Melih2011CPC, Forster1948AP}, Modified Redfield Theory \cite{NovoderezhkinVladimirI2007BJ, NovoderezhkinVladimirI2010PCCP}, and the Hierarchical Equations of Motion (HEOM) \cite{LiuHao2014JCP} for solving it, but these methods have problems such as high computational complexity. With the development of experimental techniques, two-dimensional electronic spectroscopy (2DES) has been used to observe key mechanisms in the energy transfer process of photosynthesis, including the coupling structures, energy transfer pathways, and quantum coherence within pigment-protein complexes. In the late 1980s, the molecular structure of the purple bacterial reaction center protein complex was obtained \cite{Deisenhofer1989BR}, and subsequently, several other molecular structures of reaction centers and light-harvesting antenna protein complexes were widely determined \cite{Deisenhofer199ARBBC, Mostame2012NJP, Potovcnik2018NC, TaoMingJie2020QE, Cheng2006PRL}. In 2005 \cite{BrixnerTobias2005Nature}, 2DES was used to study the energy transfer pathways in the Fenna–Matthews–Olson (FMO) photosynthetic complex. At the same time, Engel et al. \cite{ReadElizabethL2008BJ} used 2DES to study the energy transfer process in the FMO photosynthetic complex of green sulfur bacteria, revealing the key role of quantum coherence in efficient energy transfer during photosynthesis. Current research has demonstrated that quantum effects play a very important role in photosynthesis, and many researchers have also offered views on the specific roles that quantum effects play in energy transfer.

On the other hand, in animal navigation, entanglement between a pair of natural qubits (radicals) has been shown to persist for more than a few milliseconds under ambient conditions, which is much longer than the entanglement persistence time in artificial quantum systems under very low temperatures and vacuum conditions \cite{Cai2010PRL, Cai2011PRL, GaugerErikM2011PRL}. To reveal the physical mechanism of animal navigation, researchers have proposed the radical-pair mechanism (RPM) as an explanation. In the radical pair hypothesis, interconversion between the spin singlet and triplet states induced by the geomagnetic field, together with local fields caused by nuclear spins, leads to distinguishable chemical reaction products. This magnetically sensitive chemical reaction can be well described by a generalized Holstein model that includes spin degrees of freedom \cite{YangLiPing2012PRA}.

Although many methods have been proposed to study energy transfer in photosynthesis, most of these theories are based on certain approximations and have their own ranges of using and limitations. In addition, some theories have very high computational complexity, which to some extent is not helpful for in-depth study of quantum effects in photosynthesis. Due to superposition and entanglement, quantum computing can achieve true parallel computation and has advantages that classical computing cannot match. Therefore, for photosynthetic systems, one can use quantum simulation to compute the energy transfer process \cite{ZhangNaNa2021FOP, WangBiXue2018npj, ChenXinYu2022npj, Mostame2012NJP, MakriNancy1996PNAS, CalvinMelvin1978ACR, LiguoriNicoletta2020PR, Potovcnik2018NC, TaoMingJie2020QE, Gorman2018PRX, Cirac1995PRL, Haffner2008PR, Blatt2012NP}. Similarly, quantum simulation is also suitable for studies of animal navigation. At present, researchers have realized quantum simulations of the photosynthetic process and animal navigation on multiple platforms \cite{Cai2010PRL, Zhang2023JPCL, Muniz2020Nature, CaiCY22012PRA, WuJiaYi2022JPCB}.

Inspired by the roles of quantum coherence in biological systems for improving energy transfer efficiency and enhancing metrological precision, researchers have proposed theoretical and experimental frameworks for constructing quantum coherent devices. Crearore et al. \cite{Creatore2013PRL} introduced a simplified model in which two dipole-coupled pigments form “bright” and “dark” states that replace the natural reaction center. Photons first excite the bright state, which then rapidly undergoes nonradiative transfer to the dark state and injects an electron into an artificial acceptor, thereby avoiding radiative losses while increasing injection efficiency. This exciton coherence-based design can increase photocurrent and maximize power output. Xiao et al. \cite{Xiao2020PRL}, inspired by animal navigation, proposed that as long as resonance and triplet nondegeneracy conditions are met, local magnetic field fluctuations (magnetic noise) produced by magnetic proteins can drive interconversion between the singlet and triplet states of radical pairs and enhance directional sensitivity to the geomagnetic field, indicating that magnetic noise can enhance metrological precision.

This paper focuses on photosynthesis and animal navigation in quantum biology, and summarizes quantum coherence effects in these two classes of biological systems and their theoretical and experimental methods. It further summarizes how researchers have used quantum simulation across multiple platforms to realize these two processes, as well as the device-level frameworks for quantum coherent technologies inspired by them and their applications.

\section{Quantum effects in photosynthesis}

\subsection{Theoretical studies of photosynthesis}

The study of energy transfer in photosynthetic systems requires the application of quantum dissipative dynamics methods. In such investigations, the pigment molecules and other relevant structures of interest are typically treated as the system, while the remaining structures, such as the protein scaffolds, are treated as the environment. To date, various quantum dissipative dynamics methods have been proposed \cite{Melih2011CPC, Forster1948AP,  NovoderezhkinVladimirI2007BJ, NovoderezhkinVladimirI2010PCCP, SchroderMarkus2006JCP, JeskeJan2015JCP, LiuHao2014JCP}.

Theoretical approaches to studying photosynthesis include the F{\"o}rster theory, which describes stochastic resonance energy transfer between weakly coupled pigment molecules \cite{Melih2011CPC, Forster1948AP}. The Modified Redfield Theory is suitable for strongly coupled pigment molecules \cite{NovoderezhkinVladimirI2007BJ, NovoderezhkinVladimirI2010PCCP}. The Generalized Bloch-Redfield Theory characterizes non-Markovian open quantum systems \cite{SchroderMarkus2006JCP, JeskeJan2015JCP}.  The HEOM enables highly accurate simulations of energy transfer between pigment molecules \cite{LiuHao2014JCP}. However, these theoretical methods all had their own ranges of applicability, detailed discussions can be found in the review by Tao et al. \cite{TaoMingJie2020SciBull}.

\begin{figure*}
    \centering
    \includegraphics[width=18 cm]{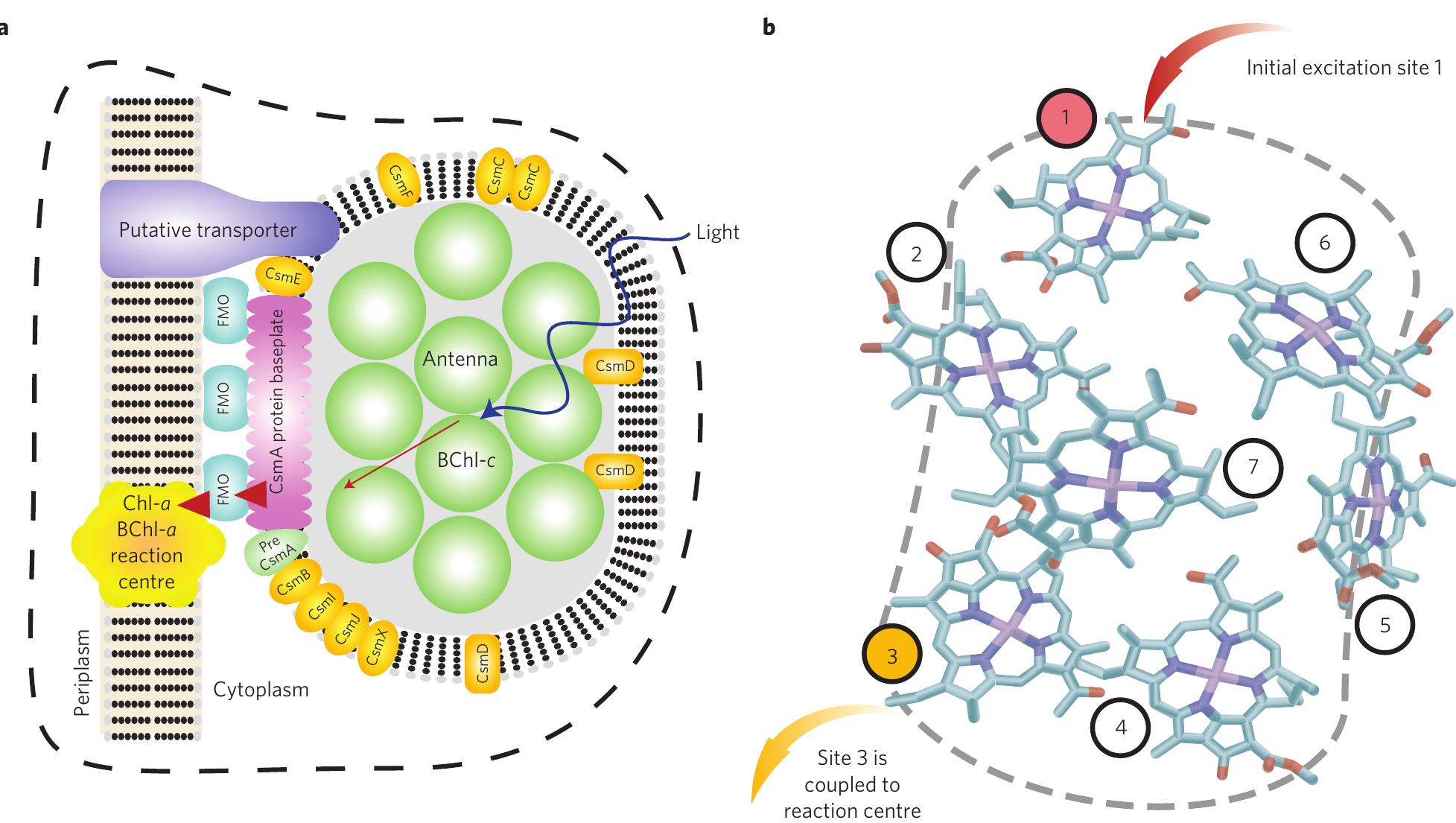}
    \caption{A quantum machine for efficient light-energy harvesting. The FMO complex, a well-known part of the light-harvesting system in green sulfur bacteria, shows signs of quantum coherent energy transfer. Both experiments and theory have examined how this protein moves energy and how quantum it really is. Work in this area could uncover quantum effects that help boost the efficiency of energy capture in living systems. \textbf{a}, Schematic of the photosynthetic system in green sulfur bacteria, comprising the antenna, the energy-transmitting baseplate, the FMO complexes, and the reaction center, the chlorosome antenna (green discs) contains about 200,000 BChl-c molecules and is an unusually large structure optimized to capture as many photons as possible under the low-light conditions preferred by these bacteria, and sunlight generates an excitation in this antenna that is passed (red arrows) to the reaction center via one of several FMO complexes. \textbf{b}, X-ray diffraction reveals the BChl-a arrangement in an FMO pigment–protein complex. The FMO complex contains eight BChl-a molecules (seven shown) embedded in a protein scaffold (not depicted). The excitation typically arrives from the chlorosome at site 1, then migrates from one BChl to the next, and upon reaching site 3 it can irreversibly transfer to the reaction center to initiate charge separation.\cite{LambertNeill2013NP}}
    \label{fig1}
\end{figure*}

\subsection{Two-dimensional electronic spectroscopy}

2DES is an ultrafast laser spectroscopy technique that offers greater resolution compared to traditional linear spectroscopy, enabling the elucidation of coherence, coupling, and dynamic processes within a system \cite{SunZongHao2023AQT, ZhangHaoYue2025JCTC, JinJingYiRan2025JCP, Weng2018CJCP, Yue2017CPL}.
Similarly, 2DES has played a pivotal role in photosynthesis research, enabling the elucidation of key mechanisms such as the coupling architecture, energy transfer pathways, and quantum coherence within pigment-protein complexes \cite{BrixnerTobias2005Nature, ReadElizabethL2008BJ, TiwariVivek2018NC, GinsbergNaomiS2009ACR,  GabrielaSSchlauCohen2011CP, LewisKristinLM2012JPCL, Zigmantas2006PANS}.

Using noncollinear three-pulse photon-echo spectroscopy, researchers have successfully recorded 2DES to investigate energy-transfer pathways in the FMO complex. Brixner et al. \cite{BrixnerTobias2005Nature} directly demonstrated the existence of two parallel energy-transfer routes, showing that energy flow in the FMO complex does not simply follow a downhill hopping mechanism. Instead, it is guided by the spatial distribution of excitonic wavefunctions, emphasizing the structural and functional complexity of energy transfer in photosynthetic systems.

Polarization-dependent 2DES has further advanced the understanding of excitonic structures and energy-transfer dynamics in the FMO complex of green sulfur bacteria. Read et al. \cite{ReadElizabethL2008BJ} showed that nonrephasing spectra provide a clearer resolution of interexciton energy-transfer pathways compared to rephasing spectra. By applying specific laser-pulse polarization combinations, their work determined projection angles between excitonic states, revealing distinct energy-transfer processes and offering detailed insights into the interplay between excitonic states.

The development of spatially resolved fluorescence-detected two-dimensional electronic spectroscopy (SF-2DES) has enabled unprecedented spatial and temporal resolution in the study of photosynthetic systems. Tiwari et al. \cite{TiwariVivek2018NC} demonstrated that this technique, which integrates femtosecond temporal resolution with submicron spatial resolution, can map the spatial heterogeneity of excitonic structures in photosynthetic bacteria. Their findings revealed that membrane remodeling under varying light-intensity growth conditions induces significant alterations in the excitonic structure of light-harvesting complexes. By analyzing intensity variations of diagonal and cross-peaks in the 2D spectra, they provided critical insights into how environmental factors influence excitonic organization and energy-transfer dynamics.

Overall, 2DES functioned as a nonlinear laser spectroscopy method with ultrafast time and frequency resolution, and it provided more accurate descriptions of coherence, coupling, and dynamics in a system than traditional linear spectroscopy, so it held a key methodological role in studies of energy transfer in photosynthesis \cite{BrixnerTobias2005Nature, ReadElizabethL2008BJ, TiwariVivek2018NC, GinsbergNaomiS2009ACR,  GabrielaSSchlauCohen2011CP, LewisKristinLM2012JPCL, Zigmantas2006PANS}. Its application to the FMO complex directly revealed parallel energy pathways and energy flow guided by the spatial distribution of exciton wave functions \cite{BrixnerTobias2005Nature}. Polarization-dependent 2DES improved the resolution of transfer pathways between exciton states by optimizing pulse polarization combinations and by using nonrephasing spectra, and it quantified projection relations and interactions between different exciton states \cite{ReadElizabethL2008BJ}. Furthermore, SF-2DES combined femtosecond time resolution with submicron spatial resolution, produced images of spatial heterogeneity in exciton structure in photosynthetic bacteria, and showed through analysis of diagonal and cross peaks in the two-dimensional spectra how membrane remodeling under different light intensities reshaped exciton organization in light-harvesting complexes and regulated energy-transfer dynamics \cite{TiwariVivek2018NC}. Together, 2DES and its variants provided fine characterization of energy coupling, coherence, and multi-path routing in photosynthesis across time, frequency, and space, and they served as essential tools for clarifying the mechanisms of light harvesting and transfer.

\subsection{Quantum coherence effects in photosynthetic systems}
Quantum coherence effects play a crucial role in photosynthetic systems. Recognizing the significance of quantum coherence, numerous studies have explored leveraging these effects to enhance EET efficiency in photosynthetic systems \cite{TaoMingJie2020SciBull, AiQing2014NJP, AiQing2013PCL, LeeHohjai2007Science, EngelGregoryS2007Nature,  LloydSeth2011JPCS, IshizakiAkihito2012ARCMP, KassalIvan2013PCL, GrahamRFleming2011PC, ChenuAurelia2015ARPC, PanitchayangkoonGitt2011PNAS, OlayaCastroAlexandra2008PRB, ScholesGregoryD2010PCL, Zhu2024NC}.

The mechanism of highly efficient energy capture and transfer in photosynthesis relies on the complex molecular structure of pigment-protein complexes. One of the simplest and most thoroughly investigated cases in the study of photosynthesis is the light-harvesting apparatus of green-sulphur bacteria, as shown in Fig. \ref{fig1}. These have a very large chlorosome antenna that allows them to thrive in low-light conditions. The energy collected by these chlorosomes is transferred to the reaction centre through a specialized structure called the FMO complex. Lee et al. \cite{LeeHohjai2007Science} investigated the role of quantum coherence in this transfer process by designing a bichromatic electronic coherent photon echo experiment. They directly probed the electronic coherence between the excitonic states of bacteriochlorophyll (BChl) and bacteriopheophytin in the reaction center of purple bacteria (Rhodobacter sphaeroides), revealing how the protein environment preserves quantum coherence and optimizes energy transfer efficiency. Experimental results demonstrated that the coherence between the H and B excitonic states lasts approximately 440 fs at low temperatures and 310 fs at higher temperatures. These coherence times significantly exceed predictions from conventional theories.
Theoretical analysis further indicated that electronic coupling alone is insufficient to account for such prolonged coherence times, necessitating consideration of nuclear motion cross-correlations in the protein environment that regulate energy fluctuations of the excitonic states. This strong correlation extends electronic coherence and enhances energy transfer efficiency. The study suggests that quantum coherence enables excitons to move rapidly and reversibly through space, efficiently locating energy traps, while the protective role of the protein environment significantly amplifies this process. 
This study showed that photosynthetic energy transfer did not follow purely incoherent F{\"o}rster hopping under the independent bath assumption, but instead included significant coherent contributions and coupling to specific vibrational modes, which made it necessary to explicitly consider long-range correlations between the system and the bath and the dynamics of coherence in the theoretical framework. This explained the efficient, fast, and reversible search and capture of energy, and it provided clear principles of coherence engineering and control of environmental correlations for the design of bioinspired solar energy conversion materials and devices.

Meanwhile, Engel et al. \cite{EngelGregoryS2007Nature} investigated the energy transfer process in the FMO photosynthetic complex of green sulfur bacteria using 2DES, uncovering the critical role of quantum coherence in efficient energy transfer during photosynthesis. For the first time, they directly observed electronic quantum coherence lasting up to 660 fs under 77 K. This phenomenon was manifested as quantum beat signals between excitons and exceeded the predictions of conventional theories regarding coherence dissipation times.
By recording the oscillatory behavior of the lowest-energy excitonic states and their cross peaks in the FMO complex, they found that these oscillations corresponded to quantum coherence between excitonic states. Their study demonstrated that quantum coherence enables the excited state to efficiently sample multiple excitonic states in a wave-like manner, thereby rapidly identifying the optimal energy transfer pathway. This mechanism significantly enhances the efficiency of energy transfer and stands in stark contrast to the traditional hopping model \cite{AbramaviciusDarius2004PCCP}.

How does quantum coherence drive energy transfer by coupling excitonic coherence with the population dynamics of excited states? Panitchayangkoon et al. \cite{PanitchayangkoonGitt2011PNAS} provided direct evidence of quantum transport in photosynthetic systems using 2DES, revealing how quantum coherence drives energy transfer through the coupling of excitonic coherence with the population dynamics of excited states. Using the FMO complex of green sulfur bacteria as a model system, they demonstrated that quantum coherence not only persists for extended periods but also directly influences the energy transfer process through oscillatory population distributions of the excited states.
The observed oscillatory behavior of the excited-state populations suggests that the protein environment in photosynthesis not only protects quantum coherence but also plays an active role in energy transfer. Through an in-depth analysis of the quantum dynamics of the FMO complex, they identified a 90° phase relationship between the population oscillations and coherence, confirming that these oscillations result from quantum transport rather than classical quantum beating or vibrational coupling. By applying the Redfield master equation theory, the authors elucidated the mechanism driving population dynamics, highlighting that the direct coupling between populations and coherence is the key to inducing these oscillations. Moreover, they found that dynamic coupling between excitonic states and the protein environment enables reversible energy exchange between the excitonic system and the protein matrix, further supporting the active role of the protein environment in quantum transport. This work clarified that the protein environment not only protected coherence in function but also took part in reversible exchange of energy with pigments through strong coupling, revealing at the microscopic level a cooperative flow of energy between excitons and bath modes. This evidence provided direct kinetic support for achieving efficient and robust photosynthetic energy transfer at physical temperatures, and it provided verifiable design guidelines for coherence engineering and environmental control in bio-inspired solar energy materials.

These studies \cite{TaoMingJie2020SciBull, AiQing2014NJP, AiQing2013PCL, LeeHohjai2007Science, EngelGregoryS2007Nature,  LloydSeth2011JPCS, IshizakiAkihito2012ARCMP, KassalIvan2013PCL, GrahamRFleming2011PC, ChenuAurelia2015ARPC, PanitchayangkoonGitt2011PNAS, OlayaCastroAlexandra2008PRB, ScholesGregoryD2010PCL} indicate that the synergy between quantum coherence and the protein environment is a core mechanism underlying the highly efficient energy transfer in photosynthetic systems. A deeper understanding of these quantum dynamical mechanisms not only expands our knowledge of natural photosynthesis but also provides critical theoretical support and experimental foundations for the design of efficient artificial light-harvesting systems.

\subsection{Applications of quantum simulation in photosynthesis}

In recent years, quantum simulation of photosynthetic systems has made notable advances on various artificial platforms, with the main aim of using controllable quantum systems to replicate the high efficiency and directionality of EET observed in natural light-harvesting complexes.
Applications of quantum simulation in photosynthesis include nuclear magnetic resonance (NMR)-based quantum simulations \cite{ZhangNaNa2021FOP, WangBiXue2018npj, ChenXinYu2022npj, Mostame2012NJP, MakriNancy1996PNAS, CalvinMelvin1978ACR, LiguoriNicoletta2020PR}, superconducting quantum circuit-based quantum simulations \cite{Potovcnik2018NC, TaoMingJie2020QE}, and ion trap-based quantum simulations \cite{Gorman2018PRX, Cirac1995PRL, Haffner2008PR, Blatt2012NP}.  

The Hamiltonian of the photosynthetic system can be expressed as
\begin{align}
H &= H_\text{S} + H_\text{B} + H_\text{SB}, \label{eq:photosynthetic_hamiltonian}
\end{align}
where  \( H_\text{S} \) represents the system Hamiltonian, \( H_\text{B} \) represents the environment Hamiltonian, \( H_\text{SB} \) represents the interaction between the system and the environment.

When simulating the quantum dynamics of the photosynthetic system on an NMR platform, the Hamiltonian of the NMR system can be expressed as
\begin{align}
H &= H_\text{S} + H_\text{PND}, \label{eq:nmr_hamiltonian}
\end{align}
where the noise Hamiltonian \( H_\text{PND} \) is given by
\begin{align}
H_\text{PND} &= \sum_m \beta_m(t)|m\rangle\langle m|, \label{eq:noise_hamiltonian}
\end{align}
and \( \beta_m(t) \) represents the time-domain distribution of the random noise. The explicit expression for \( \beta_m(t) \) is
\begin{align}
\beta_m(t) &= \sum_{j=1}^{N_c} \alpha_z^{(m)} F(\omega_j) \omega_j \cos(\omega_j t + \phi_j^{(m)}), \label{eq:noise_amplitude}
\end{align}
where \( \alpha_z^{(m)} \) is the noise amplitude acting on state \( |m\rangle \), \( \omega_j = j\omega_0 \), with \( \omega_0 \) being the base frequency, and \( N_c \omega_0 \) the cutoff frequency, \( F(\omega_j) \) is the modulation function, \( \phi_j^{(m)} \) is the random phase.
By performing the Fourier transform of the second-order correlation function of \( \beta_m(t) \), the power spectral density \( S_m(\omega) \) of the noise can be obtained as
\begin{align}
S_m(\omega) &= \int_{-\infty}^\infty d\tau e^{-i\omega\tau} \langle \beta_m(t + \tau) \beta_m(t) \rangle, \label{eq:power_spectral_density}
\end{align}
which yields
\begin{align}
S_m(\omega) &= \frac{\pi (\alpha_z^{(m)})^2}{2} \sum_{j=1}^{N_c} [\omega_j F(\omega_j)]^2 [\delta(\omega - \omega_j) + \delta(\omega + \omega_j)]. \label{eq:power_spectral_density_final}
\end{align}
Assuming the noise power spectral density \( S_m(\omega) \) has the same form as the spectral density function of the photosynthetic system, the corresponding modulation function \( F(\omega_j) \) can be determined. Subsequently, the time-domain distribution \( \beta_m(t) \) can be derived, and the noise Hamiltonian \( H_\text{PDN} \) can be constructed.

On the NMR platform, Zhang et al. \cite{ZhangNaNa2021FOP} extended a combined Hamiltonian‐ensemble, bath‐engineering, and the Gradient Ascent Pulse Engineering (GRAPE) method. They showed that an $N$‐level photosynthetic system can be encoded with just $\log_2 N$ qubits, mapping its Frenkel–exciton Hamiltonian plus any pure‐dephasing noise spectrum into random phase modulations implementable in NMR. Two- and three- qubits experiments at room temperature successfully reproduced coherent Rabi oscillations at short times and thermal equilibrium at long times. In the high-temperature limit, their results closely match numerical solutions of the HEOM. This study used quantum simulation as a tool to accurately evolve open quantum dynamics in the high temperature limit, proved that the efficiency of energy transfer was jointly optimized by a moderately clustered geometry and an environment whose spectral density peak matched the system, and showed that this held under Drude–Lorentz and other spectral densities. It provided a scalable alternative to the computational bottleneck of traditional HEOM in high dimensional and complex spectral density cases and systematically validated design rules where geometry and noise engineering together controlled the channels and rates of EET.

In 2018, Wang et al. \cite{WangBiXue2018npj} employed NMR quantum simulation to investigate EET in photosynthesis by using a linear tetramer model. The model consisted of four chromophores with site energies of 13000 cm$^{-1}$, 12900 cm$^{-1}$, 12300 cm$^{-1}$, and 12200 cm$^{-1}$, and coupling strengths of 126 cm$^{-1}$, 16 cm$^{-1}$, 132 cm$^{-1}$, and 5 cm$^{-1}$, respectively, representing typical parameters in photosynthetic systems. Environmental noise, characterized by a Drude--Lorentz spectral density, was incorporated into the simulation to study its impact on the EET process.
Experimental results demonstrated that the quantum simulation closely matched the theoretical predictions obtained using the HEOM in the short-time regime, where coherent oscillations were prominently observed. Over longer timescales, the system gradually thermalized, reaching a steady-state equilibrium, thereby validating the role of environmental dephasing in the EET process. Moreover, the study revealed that increasing the number of ensemble samples significantly reduced the discrepancy between the quantum simulation and theoretical results. For instance, while smaller sample sizes introduced noticeable differences, increasing the sample size to 500 or more led to an almost perfect agreement between the quantum simulation and HEOM calculations.
To minimize experimental errors and enhance simulation accuracy, the authors employed the GRAPE algorithm to optimize the pulse sequences. This ensured reliable results and demonstrated the potential of NMR quantum simulation to efficiently reproduce quantum coherence and environmental dephasing effects in photosynthetic energy transfer. The study provides a robust framework for exploring complex EET mechanisms and understanding the role of quantum coherence in natural photosynthetic systems.

In recent years, superconducting quantum circuits, which are highly controllable artificial quantum systems, have been used to simulate the main processes of EET in biological photosynthesis. In 2018, Poto{\v{c}}nik et al. \cite{Potovcnik2018NC} built a simple three-site model using three tunable transmon qubits and a microwave resonator. Qubits Q1 and Q2 were strongly coupled to produce bright and dark states. Qubit Q3, coupled via the Purcell effect, served as a reaction center to extract excitations. By adding controlled broadband white noise and narrowband Lorentzian noise to Q2, they showed that noise can open an energy path between the bright state and Q3 even when they are not in resonance, enabling efficient noise‐assisted energy transfer. When the noise spectrum was shaped to match the energy gap, the transfer efficiency rose to 58\% compared with 39\% under uniform noise. Their measurements also revealed the fluorescence spectra of the bright and dark states, the splitting of the intermediate doublet, and the transition between strong and weak coupling regimes. In both coherent and incoherent excitation experiments, they observed dark‐state protection and quantum interference effects. These results clearly demonstrate how quantum coherence and classical decoherence work together to guide energy transport. 

Tao et al.\cite{TaoMingJie2020QE} proposed and experimentally simulated a clustered photosynthetic model in the superconducting quantum circuit made of four charge qubits coupled to two transmission line resonators. Using a Fr$\ddot{\text{o}}$hlich–Nakajima transformation, they derived an effective Hamiltonian for the four qubits. By tuning both the qubit–resonator and the resonator–resonator couplings, they could control the coupling strengths between nearest neighbors, within each cluster, and between clusters through distance and energy level matching. Numerical simulations of the Lindblad master equation showed that when the within‐cluster coupling is increased to a moderate level, exciton delocalization and energy matching work together to maximize both the speed and the efficiency of EET. However, if the within-cluster coupling is too strong, the energy-level mismatch develops, and the transfer is suppressed. This work built and verified in a controllable superconducting quantum circuit the mechanism and optimal conditions by which a moderately clustered geometry optimized EET through exciton delocalization and energy matching, and it showed an experimental map where geometry engineering and environmental dephasing together determined the intercluster transfer rate and direction.

Trapped‐ion quantum systems are widely used to study quantum dynamics because they allow precise control and measurement of both electronic and motional degrees of freedom \cite{Gorman2018PRX, Cirac1995PRL, Haffner2008PR, Blatt2012NP}. In 2018, Gorman et al. \cite{Gorman2018PRX} employed two \(\mathrm{^{40}Ca^+}\) ions to simulate vibrationally assisted energy transfer as it occurs in photosynthesis. The ions’ electronic states are mapped to a donor and an acceptor levels. A coupling of strength \(J\) between these levels is implemented via a M{\o}lmer and S{\o}rensen interaction on the axial motion mode. The radial rocking mode, prepared in a thermal state, serves as the vibrational environment and is coupled to the donor with strength \(\kappa\) through a local two‐tone laser beam. An energy detuning \(\Delta\) between donor and acceptor is introduced by applying an optical Stark shift via unequal beam powers.
In the experiment, all motional modes are first cooled by sideband cooling to set their mean phonon occupation. The system is then initialized in the state \(\lvert DS\rangle\), evolved under the simulated Hamiltonian for a variable time \(\tau\), and finally the transfer probability is read out by fluorescence detection. They find that if \(\Delta\) exceeds \(J\), direct transfer is suppressed but the environment bridges the energy gap, producing clear single‐phonon resonance peaks in the transfer spectrum. At low temperature, the asymmetry of these peaks reveals the quantum nature of the environment. This study used a minimal and controllable model on a trapped ion quantum platform to demonstrate the key dynamical and spectroscopic features of vibrationally assisted energy transfer, revealed that when a vibrational mode matched the donor acceptor energy gap in a quantum sense the environment significantly promoted intersite energy transfer through single phonon and multiphonon processes, provided controlled causal evidence for the long hypothesized participation of specific vibrational modes and tuning of environmental spectra in photosynthesis, clarified that the environment did not only cause dephasing but could be engineered to enhance transfer, and thus laid a solid foundation for understanding and bioinspired optimization of photosynthetic energy transfer mechanisms.

\subsection{Quantum coherence devices}

Quantum simulations reveal complex energy transfer pathways and enable the optimization of artificial light-harvesting system designs, such as designing quantum coherence devices to enhance energy transfer efficiency \cite{Blankenship2021book, Blankenship2011Science, Romero2017Nature, Gratzel1991CIC, Creatore2013PRL, RomeroElisabet2014NP, AlharbiFahhadH2015RSER, BredasJeanLuc2017NM, YaoYiXuanADP2023}.

In plants and algae, the efficiency of converting photons into electrons can reach 100\% \cite{Blankenship2011Science, Blankenship2021book}.
In 2011, Blankenship et al. \cite{Blankenship2011Science} first proposed that natural photosynthesis can achieve almost 100\% quantum efficiency in its initial energy storage step. This high efficiency arises because two pigment complexes in the reaction centre absorb photons and, through a series of controlled electron transfer and energy relaxation steps, carry out highly selective charge separation. They compared the efficiency of this natural process with that of conventional silicon photovoltaics coupled to water electrolysis for hydrogen production, and suggested that, to combine high conversion efficiency with biocompatibility, one could recreate a light harvesting architecture in artificial systems.

Building on this idea, in 2013 Creatore et al. \cite{Creatore2013PRL} introduced a simplified model in which two dipole‐coupled pigments form “bright” and “dark” states that replace the natural reaction centre. Photons first excite the bright state, which then undergoes rapid nonradiative transfer to the dark state and injects electrons into an artificial acceptor, thus avoiding radiative losses while enhancing injection efficiency. Their simulations show that, compared with uncoupled pigments, this  exciton‐coherence‐based design can increase both photocurrent and maximum power output.

During the past decade, a variety of quantum simulations and spectroscopic experiments have mutually reinforced the view that electronic–vibronic coherence in natural photosynthesis can persist for picoseconds even within the noisy, highly disordered protein environment and, more importantly, can drive the ultrafast, directional separation of charge between exciton and charge‐transfer (CT) states. Romero et al.  \cite{RomeroElisabet2014NP}, studying the plant PSII reaction center, combined two‐dimensional electronic spectroscopy with Redfield‐theory simulations to show that vibrational modes at 120, 340, and 730~cm$^{-1}$ resonate with exciton–CT energy gaps, thereby sustaining and regenerating electronic coherence. Their results demonstrate that, when coherence is strong, the PD2$^{+}$–PD1$^{-}$ CT state (involving the primary chlorophyll pair) forms within tens of femtoseconds, yielding ultrafast and highly efficient charge separation; when coherence is weak, the separation slows and its efficiency drops. Further analyses using quantum master equations and Redfield models confirm a positive correlation between the strength of coherence and charge‐separation efficiency, and show that vibronic coherence helps counteract disorder‐induced decoherence. They clarified that in the PSII reaction center the electron vibrational coherence supported by specific intramolecular vibrational modes maintained and regenerated electronic coherence within the critical window of charge separation, and it showed a strong correlation with ultrafast and high yield initial charge separation, thus providing a verifiable mechanism of coherence vibration synergy that enhanced energy conversion efficiency. It provided guidance for designing quantum coherent devices for efficient energy conversion, where one achieved a match between the energy gaps of excitons and charge transfer states and the vibrational frequencies through level and coupling layouts to amplify coherence driven pathway selection and directional transport, and where one extended coherence lifetimes and established an optimum region between coherent back and dephasing by controlling the environmental spectral density and nonequilibrium vibrational states.

In the field of artificial photovoltaics, Alharbi and Kais \cite{AlharbiFahhadH2015RSER} used the detailed balance model to quantify the theoretical efficiency limits of single‐junction, multi‐junction, and split‐spectrum systems. They showed that, by borrowing the coherence‐based recombination‐suppression mechanism found in photosynthetic reaction centers, one can introduce coherent interference between exciton and CT states to weaken radiative recombination channels. Their simulations suggest that ``quantum‐coherent” solar cells could exceed the Shockley–Queisser limit by tens of percentage points. Br{\'e}das et al. \cite{BredasJeanLuc2017NM} further clarified that, in moderately disordered systems analogous to the PSII reaction center, three coherence related effects supertransfer, exchange narrowing, and biased random walks, can work together to extend exciton diffusion lengths and speed up energy transfer. They recommend that organic photovoltaic designs adopt low Stokes shift dyes, for example, phthalocyanine and squaraine, and engineer weakly coupled but controllable semi‐ordered donor–acceptor interfaces. In the design of quantum coherent devices, using low Stokes shift dyes reduced the reorganization energy and spectral broadening and reduced the gap between absorption and emission to enhance excited state delocalization and F{\"o}rster spectral overlap, which in turn increased the energy transfer rate and the diffusion length. Building weakly coupled and controllable semi ordered donor acceptor interfaces allowed both ultrafast charge separation and the suppression of back recombination and reduced the voltage loss caused by interface roughness. These points provided guidance for material selection, level and coupling layout, and for interface design and environmental spectral density engineering in quantum coherent devices.

In a more macroscopic device architecture, Yao et al. \cite{YaoYiXuanADP2023} applied the photosynthetic energy‐level gradient plus noise‐assisted directionality concept to an array of coupled resonators. Using the quantum master equation and detailed balance analyses, they showed that a frequency gradient between neighboring cavities, together with coupling to a thermal bath, can produce nonreciprocal light transmission. When the frequency difference $\Delta\omega$ is much larger than the coupling strength, the transmission contrast scales as $e^{\Delta\omega/(k_{B}T)}$, enabling high unidirectional transmission in an isothermal and noisy environment without magnetic fields or time‐dependent modulation.

These insights open a path toward quantum‐engineered approaches for future solar‐energy conversion and light‐harvesting technologies.

\section{Quantum effects in migratory avian navigation}


\subsection{Quantum coherence phenomena in migratory avian navigation}

Magnetoreception is the ability of certain migratory species to navigate using the Earth’s magnetic field \cite{Wiltschko2010JRSI, Treiber2012Nature, Wiltschko2000Naturwissenschaften, Wiltschko2002Nature, Ritz2004Nature}.
Wiltschko et al. \cite{Wiltschko2010JRSI} proposed a mechanism based on magnetite, suggesting that the magnetite in the bird’s beak generates a physical torque in the magnetic field and transmits polarity or intensity information to the brain via the trigeminal nerve.
Qin et al. \cite{Qin2016NM, Guo2021SR, Zhou2023ZR} have found the animal magnetosensory receptor gene Magnetoreceptor (MagR), and have shown that the newly found receptor MagR has formed rod-like core structures by polymerizing under a magnetic field. The photoreceptors cryptochromes have wound around the outside in a helix, forming a light–magnet coupled complex with intrinsic magnetism, which has recognized and has responded to external magnetic fields, and have proposed a complex in which MagR has linearly polymerized into a rod-like core that is helically wrapped by the photoreceptor cryptochrome, termed a biocompass.

However, the European robin’s magnetoreception system functions as an inclination compass, which is insensitive to field polarity, whereas most magnetite-based compass models respond to polarity. 
Some studies have shown that their navigation relies on photoreceptors \cite{Wiltschko2010JRSI, Wiltschko2000Naturwissenschaften}.
Ritz et al. \cite{Ritz2004Nature} exposed migratory robins to both the static geomagnetic field and a radio-frequency oscillating field tuned to the radical-pair hyperfine splitting. They found that when the oscillating field was perpendicular to the geomagnetic field, the birds lost their orientation, but when the two fields were parallel, orientation was restored, demonstrating a resonance effect. In addition, Wiltschko et al. \cite{Wiltschko2010JRSI, Wiltschko2002Nature} showed that this magnetoreception is highly sensitive to changes in field strength and frequency.

Quantum coherence phenomena in migratory bird navigation involve coherent interconversion between singlet and triplet states mediated by the nuclear spin environment. The directional dependence of geomagnetic signals influences the chemical reaction rates, while quantum phase transitions in the nuclear spin environment enhance navigational sensitivity \cite{RitzThorsten2000BJ, GaugerErikM2011PRL, Bandyopadhyay2012PRL, RitzThorsten2011PC, CaiJianming2013PRL, Hiscock2016PNAS, Marais2018JRSI, Holland2014JZ, Kattnig2016PCCP}.

\begin{figure*}
    \centering
    \includegraphics[width=18 cm]{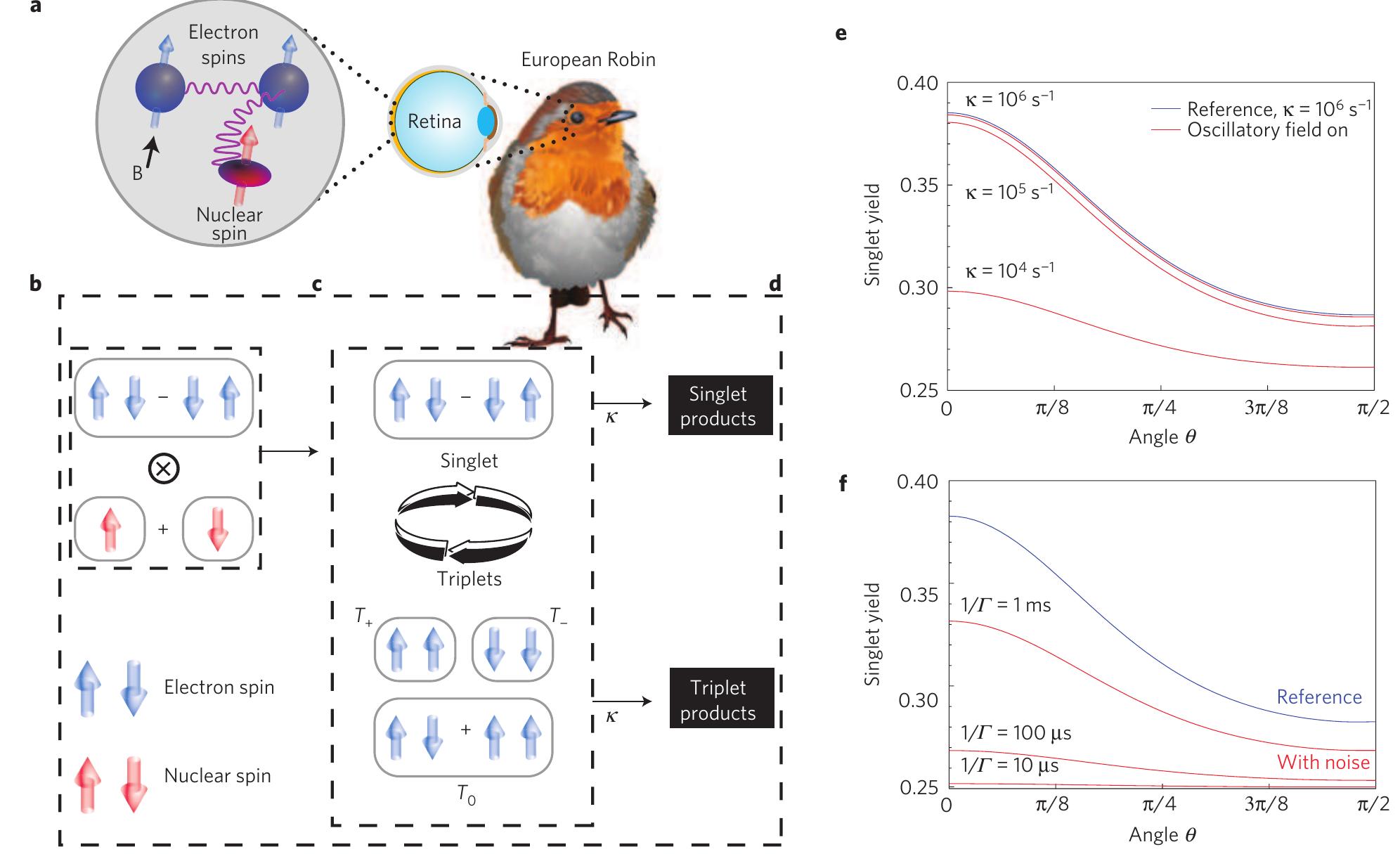}
    \caption{The avian quantum compass. The radical pair mechanism for avian magnetoreception accounted for many behavioural observations in several migrating bird species. The main features of the proposed model relied on quantum mechanics, so it may have functioned as a piece of biological quantum hardware. \textbf{a}, A diagram of the radical pair mechanism for magnetoreception that could have been used by European robins and other species. It was proposed to occur within cryptochromes in the retina and proceeded in three steps. First, light drove electron transfer from a donor to an acceptor to create a radical pair. \textbf{b, c}, Second, the singlet and triplet electron spin states interconverted due to external Zeeman and internal hyperfine magnetic couplings. \textbf{d}, Third, singlet and triplet radical pairs recombined into singlet and triplet products, which were biologically detectable. \textbf{e}, The singlet yield as a function of the external field angle $\theta$ in the presence of an oscillatory field. The blue top curve showed the yield for a static geomagnetic field with $B_0 = 47\,\mu\text{T}$, and the red curves showed the singlet yield when a $150\,\text{nT}$ field oscillating at $1.316\,\text{MHz}$ was applied perpendicular to the static field. The sensitivity of the compass was the difference in the yield between $\theta=0$ and $\theta=\pi/2$. A clear change in this sensitivity appeared once $\kappa$, the radical decay rate, reached about $10^{4}\,\text{s}^{-1}$. \textbf{f}, The singlet yield as a function of the magnetic field angle $\theta$ for different noise strengths. The blue curve showed the optimal case without noise with $\kappa=10^{4}\,\text{s}^{-1}$. The red curves showed that a general noise rate $\Gamma>0.1\kappa$ reduced the sensitivity. These results together indicated that the electron spin state required a very long coherence time. \cite{LambertNeill2013NP, GaugerErikM2011PRL}}
    \label{fig2}
\end{figure*}

Magnetoreception denotes the ability of certain migratory taxa to navigate by referring to the Earth’s magnetic field \cite{LambertNeill2013NP}.  
Schulten et al. \cite{Schulten1978ZPC} proposed the RPM to explain the light-dependent nature of magnetoreception. A schematic of the RPM is shown in Fig. \ref{fig2}.
The standard radical-pair model explains how two bonded molecules, each carrying an unpaired electron, form through a light-activated reaction and enter spin-correlated states as either singlets or triplets. In the presence of the Earth’s weak magnetic field and minor interactions with nearby atomic nuclei, these spin states evolve over time, switching between singlet and triplet configurations. Because the rate at which each radical pair recombines depends on its spin state, the process yields different chemical products. Organisms can sense these products, and since the relative proportions of singlet and triplet outcomes depend on the alignment of the external magnetic field, the pattern of reaction products conveys directional information, effectively acting as a magnetic compass.
Xie \cite{Xie2022Innovation} has identified, through structural modeling, a long-range intermolecular electron transfer chain within the MagR and cryptochrome protein complex, and has proposed that animal magnetosensing and biological navigation may be unified by integrating the radical-pair model with the biocompass via spin-dependent electron transport that couples electron spin states with intermolecular electron transfer.


There has been extensive research on the RPM \cite{GaugerErikM2011PRL, Efimova2008BJ, RitzThorsten2000BJ, Maeda2008Nature, steiner1989CR, Woodward2001PRL, Cai2012PRA, Stoneham2012BJ, Wu2012Science, YangLiPing2012PRA}. Gauger et al. \cite{GaugerErikM2011PRL} used the simplest model to describe two electron spins that started in a singlet state, where one electron had an anisotropic hyperfine coupling with a nuclear spin. The geomagnetic field caused interconversion between the singlet state and the triplet state, and spin selective recombination produced distinguishable chemical products. They computed the dependence of the singlet yield on the direction of the external magnetic field, as shown in Fig. \ref{fig2}(e), and found that a weak radio frequency field that disrupted orientation implied that the reaction rate of the radical pair did not exceed about $10^{4}\ \text{s}^{-1}$, so the coherence time needed to reach at least tens to hundreds of microseconds. As shown in Fig. \ref{fig2}(f), when the environmental noise rate exceeded about $0.1\kappa$, the angular contrast of the singlet yield dropped markedly, which emphasized that sustained long coherence and entanglement were key to maintaining the sensitivity of the singlet and triplet yields to the geomagnetic direction and thus to realizing navigation.

In recent years, many studies have focused on the sensitivity of radical pair reactions to weak magnetic fields. Woodward et al. \cite{Woodward2001PRL} studied the pyrene anion and dimethylaniline cation radical pair in solution, measuring changes in the singlet state electron hole recombination yield under radio frequency magnetic fields from 1 to 80 MHz. They found that different deuterated isotopomers produced distinct spectra and proposed a calculation method based on hyperfine coupling and perturbation theory which successfully explained the observed magnetic field and isotope effects. Maeda et al. \cite{Maeda2008Nature} carried out experiments in frozen liquid crystal and with photoselection, using a carotenoid porphyrin fullerene triad to demonstrate anisotropic changes in radical-pair lifetimes under  the geomagnetic field strength of about 50 $\mu$T. They showed that weak static magnetic fields can alter radical pair kinetics and respond selectively to field direction, providing strong evidence for the chemical compass model.

Ritz et al. \cite{RitzThorsten2000BJ} found that cryptochrome can transfer an electron to a nearby molecule. This process forms an entangled radical pair. Hyperfine interactions inside the molecule link the electron spin and nuclear spins. This makes the pair very sensitive to external magnetic fields. When the field’s direction or strength changes, the conversion rate between spin states also changes. This change alters the yield of the reaction products.
Cai et al. \cite{Cai2012PRA} used numerical and analytic methods to study how the hyperfine coupling strength of a single nuclear spin affects magnetic sensitivity. They found that a highly directional but weak coupling works best. For a realistic radical pair lifetime, this coupling yields the highest sensitivity.

There is an identifiable neural basis in pigeons for processing magnetic information. Wu et al. \cite{Wu2012Science} placed awake, head-fixed pigeons in total darkness inside a three-dimensional coil system. They then varied the magnetic field direction and strength step by step and recorded neuronal firing. Their results show that neurons in the vestibular nuclei respond directly to the geomagnetic vector. This provides experimental evidence for how birds process magnetic cues.
Earlier work proposed \cite{Wiltschko2005JCPA} that pigeon magnetoreception relied on magnetite deposits in the beak. However, a recent detailed study \cite{Treiber2012Nature} found that these deposits are actually macrophages and do not contribute to magnetoreception.
These findings do not undermine the radical-pair model. Pigeon magnetoreception is clearly polarity-sensitive, and the RPM can fully explain that feature.

\subsection{Applications of quantum simulation in migratory avian navigation}


Applications of quantum simulation in avian navigation include NMR-based quantum simulations \cite{Cai2010PRL, Zhang2023JPCL}, superconducting quantum circuit-based quantum simulations \cite{Zhang2023JPCL}, and trapped-ion and cavity-QED platform-based quantum simulations that realize critical-spin environments relevant to radical-pair dynamics \cite{Muniz2020Nature, CaiCY22012PRA, WuJiaYi2022JPCB}.

Cai et al. \cite{Cai2010PRL} performed a broad-sense quantum simulation of the RPM for avian magnetoreception. They specified a spin Hamiltonian that includes external magnetic fields and isotropic hyperfine interactions, combine it with an exponential re-encounter probability in a reaction–recombination model, and evolve the full electron–nuclear spin dynamics using a Chebyshev expansion. This allows them to compute the singlet yield $\Phi_s$, the magnetic-field sensitivity $\Lambda$, and the angular dependence $\Phi_s(\theta)$. Building on this framework, they incorporate quantum control pulses into the Hamiltonian to assess their impact. They found that periodic $\pi$ pulses along the field suppress $S$–$T$ mixing and markedly reduce sensitivity. When the field direction is periodically flipped, which degrades the compass function, synchronized $X$-axis $\pi$ pulses can restore performance. They also show that an initial $\pi/2$ pulse can distinguish an entangled singlet from a classically correlated mixed state. A comparison between Py–DMA and FADH$^{\bullet-}$–O$_2^{\bullet-}$ further indicates that, in the former, entanglement correlates with sensitivity and serves as a dynamical fingerprint, while in the latter, entanglement is short-lived and not essential. Moreover, they predict that suitable control pulses can induce angular dependence even without a static field, which suggests feasible routes for in~vitro spin chemistry or behavioral tests of the mechanism and informs the design of bioinspired weak-magnetic sensing.

Keens et al. \cite{Keens2018PRL} have shown that in a three spin system, electron dipole coupling alone has produced strong weak field effects even without nuclear hyperfine interaction. The mechanism was that dipole coupling together with an external magnetic field broke the conservation laws and symmetries present in two spin RPM systems. The field further lifted Kramers degeneracy and caused level crossings or anticrossings, which has led to strong directional anisotropy and sharp peaks in the reaction yield. These features are tunable by geometry and exchange coupling. This has shifted the understood origin of magnetosensitivity from nuclear hyperfine effects to multi spin electron interactions, and thus has provided a basis for a bionic compass that does not rely on a nuclear bath. 
Building on this idea, Cai et al. \cite{CaiCY22012PRA} showed that two central spins coupled to a critical spin bath yielded a singlet output determined by the Loschmidt echo of the environment. They demonstrated that a transverse-field Ising chain near criticality could strongly enhance the response to weak magnetic fields and their direction. Extending this criticality-based idea, Wu et al.  \cite{WuJiaYi2022JPCB} proposed a Lipkin–Meshkov–Glick (LMG) multi-spin bath to construct a bionic compass and showed that the magnetic sensitivity peaked at the LMG critical point and increased with bath size. These results generalized the enhancement mechanism from the Ising model to the LMG model and provided a concrete target Hamiltonian with measurable indicators that could be engineered. On the experimental side, Muniz et al. \cite{Muniz2020Nature} realized LMG dynamics and a tunable dynamical phase transition with nearly one million strontium atoms in an optical cavity. This platform offers a mature hardware system that can act as an LMG-type environment and, in turn, supports a testable quantum simulation scheme for bio-inspired weak magnetic sensing.

In parallel with these model-driven and hardware-ready advances, Zhang et al.  \cite{Zhang2023JPCL} focus directly on the open-system dynamics of the RPM through a general quantum algorithm. They map the Lindblad master equation into a quantum circuit by applying the Sz.-Nagy unitary dilation, which embeds nonunitary evolution into unitary operations that are implementable on a circuit. To represent spin-selective relaxation from the electronic singlet and triplet states into product states, they introduce eight projection operators and two shelving states, so that the final singlet and triplet yields follow from the occupations of the shelving states. The system Hamiltonian includes two electron spins, one nuclear spin, an anisotropic hyperfine tensor, and an external magnetic field with tunable direction and strength, with the initial state chosen as an electronic singlet and a maximally mixed nuclear spin. The algorithm applies the nonunitary part as a sequence of Kraus operators at each time step, while the unitary part is implemented by diagonalizing the Hamiltonian and applying the corresponding unitary. Each Kraus operator is realized through its unitary dilation, and projection measurements on the computational subspace at each time step yield the diagonal elements of the density matrix. These are accumulated to obtain the steady-state singlet and triplet yields and their angular dependence. Simulations on the IBM QASM simulator with geomagnetic-scale fields and typical hyperfine strengths, and with time step and decay rate chosen to reach steady state on a physical timescale, show excellent agreement with classical calculations in both angular dependence and time evolution. They applied a general open quantum dynamics algorithm to the quantum simulation of the RPM, accurately reproduced the singlet and triplet yield dynamics and the angular dependence of the avian compass model under the Lindblad framework, and thus provided a verifiable technical route for quantum computing to enter the study of biomagnetoreception.

Quantum simulation provided a testable and scalable way to study the avian navigation mechanism, reproduced in a controlled way the spin interactions and environmental effects that were hard to isolate in experiments, and accurately assessed key metrics such as the dependence on magnetic field direction and sensitivity. With programmable circuits and tunable physical platforms, it mapped open system dynamics to observables and thus enabled validation of the core processes of the chemical compass. It also laid the groundwork for designing quantum coherent devices based on light sensing.

\subsection{Quantum devices}

There were many studies that designed quantum coherent devices inspired by avian navigation\cite{Cai2010PRL, Cai2011PRL, GaugerErikM2011PRL, CaiJianming2013PRL, WuJiaYi2022JPCB, YangLiPing2012PRA, Creatore2013PRL, Xiao2020PRL, CaiCY22012PRA, Janitz2022JMCC}.
In 2010, Cai et al. \cite{Cai2010PRL} introduced quantum control into the radical pair reaction and showed that control pulses can suppress or recover the function of a chemical compass, thus providing an experimental handle to test the RPM. They computed how magnetic sensitivity depends on the initial state, compared the limits for separable and entangled states, and showed that at typical lifetime scales entanglement significantly boosted sensitivity and served as a fingerprint of the underlying spin dynamics, while for candidate systems with very long lifetimes entanglement was not required. They also presented schemes that restore angle dependence under an alternating magnetic field by phase-matched control sequences, and even induced angle readout with control alone in the absence of a static field, which highlighted quantum control as a tool for device tuning and diagnosis. This study introduced programmable pulsed quantum control and initial state engineering in a radical pair chemical compass, revealed that the cooperation of coherence and decoherence played a decisive role in weak magnetic responses, and provided a quantifiable link among entanglement, coherence lifetime, and magnetic sensitivity. It offered a practical idea for designing quantum coherent devices inspired by avian navigation, namely to turn on or suppress singlet triplet mixing as needed and to restore or induce angular dependence under alternating or zero static magnetic fields, thereby strengthening the target magnetic response and reducing background noise.

Building on this foundation, Cai \cite{Cai2011PRL} proposed a controllable strong magnetic field gradient to enhance the directional sensitivity of a chemical compass by creating a large local field at one radical while keeping the other near zero. This approach amplified directional anisotropy without reliance on complex anisotropic hyperfine couplings, and it markedly increased the singlet yield response to both the direction and the strength of the geomagnetic field. The work provided experimental designs and numerical validation in oriented liquid crystal samples and with magnetic nanoparticles, and it pointed out that the gradient also acted as a probe to distinguish spin correlations in different initial states, which enabled the detection of quantum correlations and coherence in radical pair reactions. This work showed that introducing an engineered strong local magnetic field gradient on one side of a radical pair significantly amplified the directional sensitivity of a chemical compass and linked the yield readout directly to controllable spin correlation functions, thus turning the anisotropic environment into a designable device element. This provided a biomimetic route for quantum coherent devices with hybrid metal organic architecture, used the gradient field to replace or enhance intramolecular anisotropy to improve weak magnetic responses, and enabled high sensitivity metrology.

In addition, Cai et al. \cite{CaiCY22012PRA} showed that the singlet chemical yield can be written as an integral over the Loschmidt echo of the environment, and that when each electron’s nuclear spin bath is modeled as a transverse field Ising chain the echo decayed rapidly near the quantum critical point. This behavior amplified the response of the yield to a weak external field so that the yield varied more clearly with the field direction and strength. The sensitivity was further controlled by the recombination rate, system size, temperature, and coupling strength, and the results indicated that only with interactions between nuclei did the sensitivity remained nonzero as the system size grows, which provided both an amplification mechanism and a practical parameter window. They combined the radical pair chemical compass with quantum criticality, proved that the singlet yield was determined by the Loschmidt echo of the environment, and showed that near the critical point the changes in the amplitude and direction of a weak magnetic field were strongly amplified, thus providing a biomimetic quantum coherent device paradigm that used environmental criticality as an amplifier. This provided an engineerable mechanism to enhance sensitivity by tuning the nuclear spin interaction strength, the environment size, and the recombination rate to optimize between coherence, decoherence, and critical response, thereby achieving highly sensitive detection of weak magnetic fields under feasible parameters.


Moving toward many body design, Wu et al. \cite{WuJiaYi2022JPCB} moved from two radicals to many radicals,  modeled the radical bath with the LMG model, and opened the singlet to triplet conversion by breaking central symmetry in two ways. Their numerical results showed clear changes of the singlet yield with field direction and strength and a peak of magnetosensitivity near the quantum phase transition, and as the number of radicals grew the peak became stronger and tended to converge, which indicated that many body criticality improved directional resolution and robustness. They combined multiradical systems with LMG criticality and presented a biomimetic compass paradigm of symmetry engineering and environmental critical amplification. They achieved controllable coherent interconversion by breaking microscopic symmetry and used collective criticality as a gain to ease the reliance on ultra long lifetimes and strong anisotropic hyperfine interactions, thus realizing biomimetic quantum coherent sensors with high directional resolution and robustness.


In addition, Xiao et al. \cite{Xiao2020PRL} proposed that local magnetic field fluctuations (magnetic noise) generated by magnetic proteins drive singlet–triplet interconversion of a radical pair and enhanced directional sensitivity to the geomagnetic field, provided that resonance and triplet nondegeneracy conditions were met, this suggest improved navigational precision and establish a theoretical framework for a quantum biocompass. Its results, however, have been based on short-time approximations and a spectrum–filter overlap analysis, supplemented by numerical simulations with Lindblad decoherence, and thus have constituted mechanism-oriented approximate modeling. Building on this foundation, Yang et al. \cite{Yang2020AP} incorporated the central electron spin into the clusters and introduced a generalized cluster-correlation expansion, which enabled high-accuracy, convergent calculations of population dynamics for a given Hamiltonian and initial state. Based on the magnetic noise driven interconversion mechanism of radical pairs and on high precision computational method, these works had important significance for the design of quantum coherent devices inspired by avian navigation, as long as the noise power spectrum matched the transition frequencies and the triplet degeneracy was avoided, so that high sensitivity of metrology was still maintained in the presence of decoherence.


In general, inspired by avian navigation, many studies proposed multiple methods that could enhance metrological precision and could be used to design quantum coherent devices. Quantum control adjusted on demand the interconversion between singlet and triplet in radical pair reactions and restored angle dependence, which strengthened weak magnetic responses and reduced background noise. Strong local magnetic field gradients significantly amplified directional sensitivity and turned anisotropy into a designable device element. The response of a spin bath near the quantum critical point further amplified the amplitude and directional signals of weak magnetic fields. Multiradical models opened the interconversion channel through symmetry engineering and used collective criticality as a gain, which improved directional resolution and robustness. Anisotropic magnetic noise produced by magnetic proteins also achieved direction sensitive readout when resonance and nondegeneracy conditions were satisfied and maintained high sensitivity in the presence of decoherence. These advances together formed a feasible path from microscopic mechanisms to device realization and supported the construction of highly sensitive and controllable quantum coherent sensors under realistic conditions.

\section{Conclusion}


Overall, quantum effects have broad applications in biological systems. In photosynthesis, the efficiency of energy transfer can reach 100\%. Migratory birds have a light‑dependent magnetic compass that can detect the direction of the Earth’s magnetic field with high precision. If we can effectively simulate quantum effects in biological systems and use them to design quantum coherent devices, this could enable advances in efficient energy use and high‑precision quantum sensing.

In this article, we first introduce several theories used to study photosynthesis, such as F\"orster theory and Modified Redfield theory \cite{Melih2011CPC, Forster1948AP, NovoderezhkinVladimirI2007BJ, NovoderezhkinVladimirI2010PCCP}. These methods can accurately model energy transfer between molecules. We then describe the development of 2DES. 2DES has played a key role in research on photosynthesis. It can reveal the coupling structure, energy transfer pathways, and quantum coherence in pigment protein complexes, and thus provides experimental tools for studying quantum effects in photosynthesis. Using 2DES, researchers have shown how the protein environment preserves quantum coherence and optimizes energy transfer efficiency. They have observed electronic quantum coherence and revealed its key role in efficient energy transfer during photosynthesis. These theoretical and experimental advances lay the groundwork for artificial high efficiency solar energy use and for the development of quantum coherent devices.

At the same time, due to the low complexity of quantum simulations, there have been simulations of photosynthesis based on NMR, superconducting platforms, and ion trap-based systems. Across these platforms, quantum simulations of photosynthetic systems have found coherent oscillations that agree with exact solutions such as HEOM, while the complexity is markedly reduced. Inspired by these findings, researchers have begun to use quantum coherence to build devices that achieve efficient energy transfer. Creatore et al. \cite{Creatore2013PRL} showed that a design that uses exciton coherence can increase photocurrent and maximum power output compared with uncoupled pigments. Alharbi and Kais \cite{AlharbiFahhadH2015RSER} used a coherence-based recombination suppression mechanism from the photosynthetic reaction center. Their simulations indicate that “quantum coherent” solar cells could exceed the Shockley–Queisser limit by several tens of percent. Br\'edas et al. \cite{BredasJeanLuc2017NM} suggested using low Stokes shift dyes, such as phthalocyanines and squaraines, in organic photovoltaics. In addition, Yao et al. \cite{YaoYiXuanADP2023} applied the concept of a photosynthetic energy level gradient together with noise assisted directionality to coupled resonator arrays, achieving highly unidirectional energy transfer in an isothermal and noisy environment. A growing body of theory and experiment shows that introducing quantum coherence into artificial systems can greatly enhance energy transfer efficiency, opening a path to quantum engineering methods for solar energy conversion and light harvesting technologies.

On the other hand, we next introduce quantum coherence in animal navigation that uses the geomagnetic field. Researchers usually explain animal navigation with mechanisms based on magnetite, magnetoreceptor genes, and cryptochromes. Schulten et al. \cite{Schulten1978ZPC} proposed the RPM to explain the light dependent nature of magnetoreception. Radical pairs created by photoexcitation are initially in a quantum superposition of singlet and triplet states, and the spins of the two unpaired electrons are quantum coherently coupled. The geomagnetic field modulates singlet to triplet coherent oscillations and interconversion rates through the Zeeman interaction and anisotropic hyperfine interactions.

Similarly, the low complexity of quantum simulation can be applied to animal navigation. At present, platforms based on NMR, superconducting devices, and cavity QED have all implemented the dynamics of radical pairs. Inspired by animal navigation, researchers use the RPM to build high precision chemical compasses. Cai \cite{Cai2011PRL} proposed a controllable strong magnetic field gradient that enhances directional sensitivity by creating a large local field at one site while keeping the other near zero. Wu et al. \cite{WuJiaYi2022JPCB} modeled a bath of radicals with the LMG model and enabled singlet to triplet conversion by breaking central symmetry in two ways, their results show that many body criticality improves directional resolution and robustness. Xiao et al. \cite{Xiao2020PRL} showed that, as long as resonance and triplet nondegeneracy hold, local magnetic field fluctuations generated by magnetic proteins, that is, magnetic noise, can drive singlet to triplet interconversion in radical pairs and enhance sensitivity to the geomagnetic field, indicating that magnetic noise can increase metrological precision. 

In summary, more and more studies showed that applying quantum simulation to biological systems helped to study quantum coherence effects in them more quickly and accurately. Quantum coherent devices inspired by energy transfer in photosynthesis and by the mechanism of avian navigation provided a feasible path for efficient use of energy and for achieving high precision metrology.

\section{Acknowledgments.}
This work is supported by Quantum Science and Technology-National Science and Technology Major Project(2023ZD0300200), the National Natural Science Foundation of China under Grant No.~62461160263, the National Natural Science Foundation of China under Grant No.~62131002, the Beijing Advanced Innovation Center for Future Chip (ICFC), and Tsinghua University Initiative Scientific Research Program.

\providecommand{\noopsort}[1]{}\providecommand{\singleletter}[1]{#1}%


\begin{thebibliography}{100}%
	\makeatletter
	\providecommand \@ifxundefined [1]{%
		\@ifx{#1\undefined}
	}%
	\providecommand \@ifnum [1]{%
		\ifnum #1\expandafter \@firstoftwo
		\else \expandafter \@secondoftwo
		\fi
	}%
	\providecommand \@ifx [1]{%
		\ifx #1\expandafter \@firstoftwo
		\else \expandafter \@secondoftwo
		\fi
	}%
	\providecommand \natexlab [1]{#1}%
	\providecommand \enquote  [1]{``#1''}%
	\providecommand \bibnamefont  [1]{#1}%
	\providecommand \bibfnamefont [1]{#1}%
	\providecommand \citenamefont [1]{#1}%
	\providecommand \href@noop [0]{\@secondoftwo}%
	\providecommand \href [0]{\begingroup \@sanitize@url \@href}%
	\providecommand \@href[1]{\@@startlink{#1}\@@href}%
	\providecommand \@@href[1]{\endgroup#1\@@endlink}%
	\providecommand \@sanitize@url [0]{\catcode `\\12\catcode `\$12\catcode
		`\&12\catcode `\#12\catcode `\^12\catcode `\_12\catcode `\%12\relax}%
	\providecommand \@@startlink[1]{}%
	\providecommand \@@endlink[0]{}%
	\providecommand \url  [0]{\begingroup\@sanitize@url \@url }%
	\providecommand \@url [1]{\endgroup\@href {#1}{\urlprefix }}%
	\providecommand \urlprefix  [0]{URL }%
	\providecommand \Eprint [0]{\href }%
	\providecommand \doibase [0]{https://doi.org/}%
	\providecommand \selectlanguage [0]{\@gobble}%
	\providecommand \bibinfo  [0]{\@secondoftwo}%
	\providecommand \bibfield  [0]{\@secondoftwo}%
	\providecommand \translation [1]{[#1]}%
	\providecommand \BibitemOpen [0]{}%
	\providecommand \bibitemStop [0]{}%
	\providecommand \bibitemNoStop [0]{.\EOS\space}%
	\providecommand \EOS [0]{\spacefactor3000\relax}%
	\providecommand \BibitemShut  [1]{\csname bibitem#1\endcsname}%
	\let\auto@bib@innerbib\@empty
	\bibitem [{\citenamefont {Cao}\ \emph {et~al.}(2020)\citenamefont {Cao},
		\citenamefont {Cogdell}, \citenamefont {Coker}, \citenamefont {Duan},
		\citenamefont {Hauer}, \citenamefont {Kleinekath{\"o}fer}, \citenamefont
		{Jansen}, \citenamefont {Man{\v{c}}al}, \citenamefont {Miller}, \citenamefont
		{Ogilvie} \emph {et~al.}}]{CaoJianshu2020SA}%
	\BibitemOpen
	\bibfield  {author} {\bibinfo {author} {\bibfnamefont {J.}~\bibnamefont
			{Cao}}, \bibinfo {author} {\bibfnamefont {R.~J.}\ \bibnamefont {Cogdell}},
		\bibinfo {author} {\bibfnamefont {D.~F.}\ \bibnamefont {Coker}}, \bibinfo
		{author} {\bibfnamefont {H.-G.}\ \bibnamefont {Duan}}, \bibinfo {author}
		{\bibfnamefont {J.}~\bibnamefont {Hauer}}, \bibinfo {author} {\bibfnamefont
			{U.}~\bibnamefont {Kleinekath{\"o}fer}}, \bibinfo {author} {\bibfnamefont
			{T.~L.}\ \bibnamefont {Jansen}}, \bibinfo {author} {\bibfnamefont
			{T.}~\bibnamefont {Man{\v{c}}al}}, \bibinfo {author} {\bibfnamefont {R.~D.}\
			\bibnamefont {Miller}}, \bibinfo {author} {\bibfnamefont {J.~P.}\
			\bibnamefont {Ogilvie}}, \emph {et~al.},\ }\bibfield  {title} {\enquote
		{\bibinfo {title} {Quantum biology revisited},}\ }\href
	{https://doi.org/10.1126/sciadv.aaz4888} {\bibfield  {journal} {\bibinfo
			{journal} {Sci. Adv.}\ }\textbf {\bibinfo {volume} {6}},\ \bibinfo {pages}
		{eaaz4888} (\bibinfo {year} {2020})}\BibitemShut {NoStop}%
	\bibitem [{\citenamefont {Lambert}\ \emph {et~al.}(2013)\citenamefont
		{Lambert}, \citenamefont {Chen}, \citenamefont {Cheng}, \citenamefont {Li},
		\citenamefont {Chen},\ and\ \citenamefont {Nori}}]{LambertNeill2013NP}%
	\BibitemOpen
	\bibfield  {author} {\bibinfo {author} {\bibfnamefont {N.}~\bibnamefont
			{Lambert}}, \bibinfo {author} {\bibfnamefont {Y.-N.}\ \bibnamefont {Chen}},
		\bibinfo {author} {\bibfnamefont {Y.-C.}\ \bibnamefont {Cheng}}, \bibinfo
		{author} {\bibfnamefont {C.-M.}\ \bibnamefont {Li}}, \bibinfo {author}
		{\bibfnamefont {G.-Y.}\ \bibnamefont {Chen}},\ and\ \bibinfo {author}
		{\bibfnamefont {F.}~\bibnamefont {Nori}},\ }\bibfield  {title} {\enquote
		{\bibinfo {title} {Quantum biology},}\ }\href
	{https://doi.org/10.1038/nphys2474} {\bibfield  {journal} {\bibinfo
			{journal} {Nat. Phys.}\ }\textbf {\bibinfo {volume} {9}},\ \bibinfo {pages}
		{10--18} (\bibinfo {year} {2013})}\BibitemShut {NoStop}%
	\bibitem [{\citenamefont {Fleming}\ and\ \citenamefont
		{Grondelle}(1994)}]{Fleming1994PT}%
	\BibitemOpen
	\bibfield  {author} {\bibinfo {author} {\bibfnamefont {G.~R.}\ \bibnamefont
			{Fleming}}\ and\ \bibinfo {author} {\bibfnamefont {R.~v.}\ \bibnamefont
			{Grondelle}},\ }\bibfield  {title} {\enquote {\bibinfo {title} {The primary
				steps of photosynthesis},}\ }\href {https://doi.org/10.1063/1.881413}
	{\bibfield  {journal} {\bibinfo  {journal} {Phys. Today}\ }\textbf {\bibinfo
			{volume} {47}},\ \bibinfo {pages} {48--55} (\bibinfo {year}
		{1994})}\BibitemShut {NoStop}%
	\bibitem [{\citenamefont {{\c{S}}ener}\ \emph {et~al.}(2011)\citenamefont
		{{\c{S}}ener}, \citenamefont {Str{\"u}mpfer}, \citenamefont {Hsin},
		\citenamefont {Chandler}, \citenamefont {Scheuring}, \citenamefont {Hunter},\
		and\ \citenamefont {Schulten}}]{Melih2011CPC}%
	\BibitemOpen
	\bibfield  {author} {\bibinfo {author} {\bibfnamefont {M.}~\bibnamefont
			{{\c{S}}ener}}, \bibinfo {author} {\bibfnamefont {J.}~\bibnamefont
			{Str{\"u}mpfer}}, \bibinfo {author} {\bibfnamefont {J.}~\bibnamefont {Hsin}},
		\bibinfo {author} {\bibfnamefont {D.}~\bibnamefont {Chandler}}, \bibinfo
		{author} {\bibfnamefont {S.}~\bibnamefont {Scheuring}}, \bibinfo {author}
		{\bibfnamefont {C.~N.}\ \bibnamefont {Hunter}},\ and\ \bibinfo {author}
		{\bibfnamefont {K.}~\bibnamefont {Schulten}},\ }\bibfield  {title} {\enquote
		{\bibinfo {title} {F{\"o}rster energy transfer theory as reflected in the
				structures of photosynthetic light-harvesting systems},}\ }\href
	{https://doi.org/10.1002/cphc.201000944} {\bibfield  {journal} {\bibinfo
			{journal} {ChemPhysChem}\ }\textbf {\bibinfo {volume} {12}},\ \bibinfo
		{pages} {518--531} (\bibinfo {year} {2011})}\BibitemShut {NoStop}%
	\bibitem [{\citenamefont {F{\"o}rster}(1948)}]{Forster1948AP}%
	\BibitemOpen
	\bibfield  {author} {\bibinfo {author} {\bibfnamefont {T.}~\bibnamefont
			{F{\"o}rster}},\ }\bibfield  {title} {\enquote {\bibinfo {title}
			{Zwischenmolekulare energiewanderung und fluoreszenz},}\ }\href
	{https://doi.org/https://doi.org/10.1002/andp.19484370105} {\bibfield
		{journal} {\bibinfo  {journal} {Ann. Phys.}\ }\textbf {\bibinfo {volume}
			{437}},\ \bibinfo {pages} {55--75} (\bibinfo {year} {1948})}\BibitemShut
	{NoStop}%
	\bibitem [{\citenamefont {Novoderezhkin}, \citenamefont {Dekker},\ and\
		\citenamefont {Van~Grondelle}(2007)}]{NovoderezhkinVladimirI2007BJ}%
	\BibitemOpen
	\bibfield  {author} {\bibinfo {author} {\bibfnamefont {V.~I.}\ \bibnamefont
			{Novoderezhkin}}, \bibinfo {author} {\bibfnamefont {J.~P.}\ \bibnamefont
			{Dekker}},\ and\ \bibinfo {author} {\bibfnamefont {R.}~\bibnamefont
			{Van~Grondelle}},\ }\bibfield  {title} {\enquote {\bibinfo {title} {Mixing of
				exciton and charge-transfer states in photosystem ii reaction centers:
				{M}odeling of stark spectra with modified redfield theory},}\ }\href
	{https://doi.org/10.1529/biophysj.106.096867} {\bibfield  {journal} {\bibinfo
			{journal} {Biophys. J.}\ }\textbf {\bibinfo {volume} {93}},\ \bibinfo
		{pages} {1293--1311} (\bibinfo {year} {2007})}\BibitemShut {NoStop}%
	\bibitem [{\citenamefont {Novoderezhkin}\ and\ \citenamefont {van
			Grondelle}(2010)}]{NovoderezhkinVladimirI2010PCCP}%
	\BibitemOpen
	\bibfield  {author} {\bibinfo {author} {\bibfnamefont {V.~I.}\ \bibnamefont
			{Novoderezhkin}}\ and\ \bibinfo {author} {\bibfnamefont {R.}~\bibnamefont
			{van Grondelle}},\ }\bibfield  {title} {\enquote {\bibinfo {title} {Physical
				origins and models of energy transfer in photosynthetic light-harvesting},}\
	}\href {https://doi.org/10.1039/c003025b} {\bibfield  {journal} {\bibinfo
			{journal} {Phys. Chem. Chem. Phys.}\ }\textbf {\bibinfo {volume} {12}},\
		\bibinfo {pages} {7352--7365} (\bibinfo {year} {2010})}\BibitemShut {NoStop}%
	\bibitem [{\citenamefont {Schr{\"o}der}, \citenamefont {Kleinekath{\"o}fer},\
		and\ \citenamefont {Schreiber}(2006)}]{SchroderMarkus2006JCP}%
	\BibitemOpen
	\bibfield  {author} {\bibinfo {author} {\bibfnamefont {M.}~\bibnamefont
			{Schr{\"o}der}}, \bibinfo {author} {\bibfnamefont {U.}~\bibnamefont
			{Kleinekath{\"o}fer}},\ and\ \bibinfo {author} {\bibfnamefont
			{M.}~\bibnamefont {Schreiber}},\ }\bibfield  {title} {\enquote {\bibinfo
			{title} {Calculation of absorption spectra for light-harvesting systems using
				non-{M}arkovian approaches as well as modified redfield theory},}\ }\href
	{https://doi.org/10.1063/1.2171188} {\bibfield  {journal} {\bibinfo
			{journal} {J. Chem. Phys.}\ }\textbf {\bibinfo {volume} {124}},\ \bibinfo
		{pages} {084903} (\bibinfo {year} {2006})}\BibitemShut {NoStop}%
	\bibitem [{\citenamefont {Jeske}\ \emph {et~al.}(2015)\citenamefont {Jeske},
		\citenamefont {David}, \citenamefont {Plenio}, \citenamefont {Huelga},\ and\
		\citenamefont {Cole}}]{JeskeJan2015JCP}%
	\BibitemOpen
	\bibfield  {author} {\bibinfo {author} {\bibfnamefont {J.}~\bibnamefont
			{Jeske}}, \bibinfo {author} {\bibfnamefont {J.}~\bibnamefont {David}},
		\bibinfo {author} {\bibfnamefont {M.~B.}\ \bibnamefont {Plenio}}, \bibinfo
		{author} {\bibfnamefont {S.~F.}\ \bibnamefont {Huelga}},\ and\ \bibinfo
		{author} {\bibfnamefont {J.~H.}\ \bibnamefont {Cole}},\ }\bibfield  {title}
	{\enquote {\bibinfo {title} {Bloch-{R}edfield equations for modeling
				light-harvesting complexes},}\ }\href {https://doi.org/10.1063/1.4907370}
	{\bibfield  {journal} {\bibinfo  {journal} {J. Chem. Phys.}\ }\textbf
		{\bibinfo {volume} {142}},\ \bibinfo {pages} {064104} (\bibinfo {year}
		{2015})}\BibitemShut {NoStop}%
	\bibitem [{\citenamefont {Liu}\ \emph {et~al.}(2014)\citenamefont {Liu},
		\citenamefont {Zhu}, \citenamefont {Bai},\ and\ \citenamefont
		{Shi}}]{LiuHao2014JCP}%
	\BibitemOpen
	\bibfield  {author} {\bibinfo {author} {\bibfnamefont {H.}~\bibnamefont
			{Liu}}, \bibinfo {author} {\bibfnamefont {L.}~\bibnamefont {Zhu}}, \bibinfo
		{author} {\bibfnamefont {S.}~\bibnamefont {Bai}},\ and\ \bibinfo {author}
		{\bibfnamefont {Q.}~\bibnamefont {Shi}},\ }\bibfield  {title} {\enquote
		{\bibinfo {title} {Reduced quantum dynamics with arbitrary bath spectral
				densities: {H}ierarchical equations of motion based on several different bath
				decomposition schemes},}\ }\href {https://doi.org/10.1063/1.4870035}
	{\bibfield  {journal} {\bibinfo  {journal} {J. Chem. Phys.}\ }\textbf
		{\bibinfo {volume} {140}},\ \bibinfo {pages} {134106} (\bibinfo {year}
		{2014})}\BibitemShut {NoStop}%
	\bibitem [{\citenamefont {Deisenhofer}\ and\ \citenamefont
		{Michel}(1989)}]{Deisenhofer1989BR}%
	\BibitemOpen
	\bibfield  {author} {\bibinfo {author} {\bibfnamefont {J.}~\bibnamefont
			{Deisenhofer}}\ and\ \bibinfo {author} {\bibfnamefont {H.}~\bibnamefont
			{Michel}},\ }\bibfield  {title} {\enquote {\bibinfo {title} {The
				photosynthetic reaction centre from the purple bacterium rhodopseudomonas
				viridis.}}\ }\href {https://doi.org/10.1007/BF01117044} {\bibfield  {journal}
		{\bibinfo  {journal} {Biosci. Rep.}\ }\textbf {\bibinfo {volume} {8}},\
		\bibinfo {pages} {2149--2170} (\bibinfo {year} {1989})}\BibitemShut {NoStop}%
	\bibitem [{\citenamefont {Deisenhofer}\ and\ \citenamefont
		{Michel}(1991)}]{Deisenhofer199ARBBC}%
	\BibitemOpen
	\bibfield  {author} {\bibinfo {author} {\bibfnamefont {J.}~\bibnamefont
			{Deisenhofer}}\ and\ \bibinfo {author} {\bibfnamefont {H.}~\bibnamefont
			{Michel}},\ }\bibfield  {title} {\enquote {\bibinfo {title} {High-resolution
				structures of photosynthetic reaction centers},}\ }\href
	{https://doi.org/https://doi.org/10.1146/annurev.bb.20.060191.001335}
	{\bibfield  {journal} {\bibinfo  {journal} {Annu. Rev. Biophys. Biophys.
				Chem.}\ }\textbf {\bibinfo {volume} {20}},\ \bibinfo {pages} {247--266}
		(\bibinfo {year} {1991})}\BibitemShut {NoStop}%
	\bibitem [{\citenamefont {Mostame}\ \emph {et~al.}(2012)\citenamefont
		{Mostame}, \citenamefont {Rebentrost}, \citenamefont {Eisfeld}, \citenamefont
		{Kerman}, \citenamefont {Tsomokos},\ and\ \citenamefont
		{Aspuru-Guzik}}]{Mostame2012NJP}%
	\BibitemOpen
	\bibfield  {author} {\bibinfo {author} {\bibfnamefont {S.}~\bibnamefont
			{Mostame}}, \bibinfo {author} {\bibfnamefont {P.}~\bibnamefont {Rebentrost}},
		\bibinfo {author} {\bibfnamefont {A.}~\bibnamefont {Eisfeld}}, \bibinfo
		{author} {\bibfnamefont {A.~J.}\ \bibnamefont {Kerman}}, \bibinfo {author}
		{\bibfnamefont {D.~I.}\ \bibnamefont {Tsomokos}},\ and\ \bibinfo {author}
		{\bibfnamefont {A.}~\bibnamefont {Aspuru-Guzik}},\ }\bibfield  {title}
	{\enquote {\bibinfo {title} {Quantum simulator of an open quantum system
				using superconducting qubits: {E}xciton transport in photosynthetic
				complexes},}\ }\href {https://doi.org/10.1088/1367-2630/14/10/105013}
	{\bibfield  {journal} {\bibinfo  {journal} {New J. Phys.}\ }\textbf {\bibinfo
			{volume} {14}},\ \bibinfo {pages} {105013} (\bibinfo {year}
		{2012})}\BibitemShut {NoStop}%
	\bibitem [{\citenamefont {Poto{\v{c}}nik}\ \emph {et~al.}(2018)\citenamefont
		{Poto{\v{c}}nik}, \citenamefont {Bargerbos}, \citenamefont {Schr{\"o}der},
		\citenamefont {Khan}, \citenamefont {Collodo}, \citenamefont {Gasparinetti},
		\citenamefont {Salath{\'e}}, \citenamefont {Creatore}, \citenamefont
		{Eichler}, \citenamefont {T{\"u}reci} \emph {et~al.}}]{Potovcnik2018NC}%
	\BibitemOpen
	\bibfield  {author} {\bibinfo {author} {\bibfnamefont {A.}~\bibnamefont
			{Poto{\v{c}}nik}}, \bibinfo {author} {\bibfnamefont {A.}~\bibnamefont
			{Bargerbos}}, \bibinfo {author} {\bibfnamefont {F.~A.}\ \bibnamefont
			{Schr{\"o}der}}, \bibinfo {author} {\bibfnamefont {S.~A.}\ \bibnamefont
			{Khan}}, \bibinfo {author} {\bibfnamefont {M.~C.}\ \bibnamefont {Collodo}},
		\bibinfo {author} {\bibfnamefont {S.}~\bibnamefont {Gasparinetti}}, \bibinfo
		{author} {\bibfnamefont {Y.}~\bibnamefont {Salath{\'e}}}, \bibinfo {author}
		{\bibfnamefont {C.}~\bibnamefont {Creatore}}, \bibinfo {author}
		{\bibfnamefont {C.}~\bibnamefont {Eichler}}, \bibinfo {author} {\bibfnamefont
			{H.~E.}\ \bibnamefont {T{\"u}reci}}, \emph {et~al.},\ }\bibfield  {title}
	{\enquote {\bibinfo {title} {Studying light-harvesting models with
				superconducting circuits},}\ }\href
	{https://doi.org/https://doi.org/10.1038/s41467-018-03312-x} {\bibfield
		{journal} {\bibinfo  {journal} {Nat. Commun.}\ }\textbf {\bibinfo {volume}
			{9}},\ \bibinfo {pages} {904} (\bibinfo {year} {2018})}\BibitemShut {NoStop}%
	\bibitem [{\citenamefont {Tao}\ \emph {et~al.}(2020{\natexlab{a}})\citenamefont
		{Tao}, \citenamefont {Hua}, \citenamefont {Zhang}, \citenamefont {He},
		\citenamefont {Ai},\ and\ \citenamefont {Deng}}]{TaoMingJie2020QE}%
	\BibitemOpen
	\bibfield  {author} {\bibinfo {author} {\bibfnamefont {M.-J.}\ \bibnamefont
			{Tao}}, \bibinfo {author} {\bibfnamefont {M.}~\bibnamefont {Hua}}, \bibinfo
		{author} {\bibfnamefont {N.-N.}\ \bibnamefont {Zhang}}, \bibinfo {author}
		{\bibfnamefont {W.-T.}\ \bibnamefont {He}}, \bibinfo {author} {\bibfnamefont
			{Q.}~\bibnamefont {Ai}},\ and\ \bibinfo {author} {\bibfnamefont {F.-G.}\
			\bibnamefont {Deng}},\ }\bibfield  {title} {\enquote {\bibinfo {title}
			{Quantum simulation of clustered photosynthetic light harvesting in a
				superconducting quantum circuit},}\ }\href {https://doi.org/10.1002/que2.53}
	{\bibfield  {journal} {\bibinfo  {journal} {Quantum Eng.}\ }\textbf {\bibinfo
			{volume} {2}},\ \bibinfo {pages} {e53} (\bibinfo {year}
		{2020}{\natexlab{a}})}\BibitemShut {NoStop}%
	\bibitem [{\citenamefont {Cheng}\ and\ \citenamefont
		{Silbey}(2006)}]{Cheng2006PRL}%
	\BibitemOpen
	\bibfield  {author} {\bibinfo {author} {\bibfnamefont {Y.~C.}\ \bibnamefont
			{Cheng}}\ and\ \bibinfo {author} {\bibfnamefont {R.~J.}\ \bibnamefont
			{Silbey}},\ }\bibfield  {title} {\enquote {\bibinfo {title} {Coherence in the
				{B}800 ring of purple bacteria {LH}2},}\ }\href
	{https://doi.org/10.1103/PhysRevLett.96.028103} {\bibfield  {journal}
		{\bibinfo  {journal} {Phys. Rev. Lett.}\ }\textbf {\bibinfo {volume} {96}},\
		\bibinfo {pages} {028103} (\bibinfo {year} {2006})}\BibitemShut {NoStop}%
	\bibitem [{\citenamefont {Brixner}\ \emph {et~al.}(2005)\citenamefont
		{Brixner}, \citenamefont {Stenger}, \citenamefont {Vaswani}, \citenamefont
		{Cho}, \citenamefont {Blankenship},\ and\ \citenamefont
		{Fleming}}]{BrixnerTobias2005Nature}%
	\BibitemOpen
	\bibfield  {author} {\bibinfo {author} {\bibfnamefont {T.}~\bibnamefont
			{Brixner}}, \bibinfo {author} {\bibfnamefont {J.}~\bibnamefont {Stenger}},
		\bibinfo {author} {\bibfnamefont {H.~M.}\ \bibnamefont {Vaswani}}, \bibinfo
		{author} {\bibfnamefont {M.}~\bibnamefont {Cho}}, \bibinfo {author}
		{\bibfnamefont {R.~E.}\ \bibnamefont {Blankenship}},\ and\ \bibinfo {author}
		{\bibfnamefont {G.~R.}\ \bibnamefont {Fleming}},\ }\bibfield  {title}
	{\enquote {\bibinfo {title} {Two-dimensional spectroscopy of electronic
				couplings in photosynthesis},}\ }\href
	{https://doi.org/https://doi.org/10.1038/nature03429} {\bibfield  {journal}
		{\bibinfo  {journal} {Nature}\ }\textbf {\bibinfo {volume} {434}},\ \bibinfo
		{pages} {625--628} (\bibinfo {year} {2005})}\BibitemShut {NoStop}%
	\bibitem [{\citenamefont {Read}\ \emph {et~al.}(2008)\citenamefont {Read},
		\citenamefont {Schlau-Cohen}, \citenamefont {Engel}, \citenamefont {Wen},
		\citenamefont {Blankenship},\ and\ \citenamefont
		{Fleming}}]{ReadElizabethL2008BJ}%
	\BibitemOpen
	\bibfield  {author} {\bibinfo {author} {\bibfnamefont {E.~L.}\ \bibnamefont
			{Read}}, \bibinfo {author} {\bibfnamefont {G.~S.}\ \bibnamefont
			{Schlau-Cohen}}, \bibinfo {author} {\bibfnamefont {G.~S.}\ \bibnamefont
			{Engel}}, \bibinfo {author} {\bibfnamefont {J.}~\bibnamefont {Wen}}, \bibinfo
		{author} {\bibfnamefont {R.~E.}\ \bibnamefont {Blankenship}},\ and\ \bibinfo
		{author} {\bibfnamefont {G.~R.}\ \bibnamefont {Fleming}},\ }\bibfield
	{title} {\enquote {\bibinfo {title} {Visualization of excitonic structure in
				the {F}enna-{M}atthews-{O}lson photosynthetic complex by
				polarization-dependent two-dimensional electronic spectroscopy},}\ }\href
	{https://doi.org/10.1529/biophysj.107.128199} {\bibfield  {journal} {\bibinfo
			{journal} {Biophys. J.}\ }\textbf {\bibinfo {volume} {95}},\ \bibinfo
		{pages} {847--856} (\bibinfo {year} {2008})}\BibitemShut {NoStop}%
	\bibitem [{\citenamefont {Cai}, \citenamefont {Guerreschi},\ and\ \citenamefont
		{Briegel}(2010)}]{Cai2010PRL}%
	\BibitemOpen
	\bibfield  {author} {\bibinfo {author} {\bibfnamefont {J.}~\bibnamefont
			{Cai}}, \bibinfo {author} {\bibfnamefont {G.~G.}\ \bibnamefont
			{Guerreschi}},\ and\ \bibinfo {author} {\bibfnamefont {H.~J.}\ \bibnamefont
			{Briegel}},\ }\bibfield  {title} {\enquote {\bibinfo {title} {Quantum control
				and entanglement in a chemical compass},}\ }\href
	{https://doi.org/10.1103/PhysRevLett.104.220502} {\bibfield  {journal}
		{\bibinfo  {journal} {Phys. Rev. Lett.}\ }\textbf {\bibinfo {volume} {104}},\
		\bibinfo {pages} {220502} (\bibinfo {year} {2010})}\BibitemShut {NoStop}%
	\bibitem [{\citenamefont {Cai}(2011)}]{Cai2011PRL}%
	\BibitemOpen
	\bibfield  {author} {\bibinfo {author} {\bibfnamefont {J.}~\bibnamefont
			{Cai}},\ }\bibfield  {title} {\enquote {\bibinfo {title} {Quantum probe and
				design for a chemical compass with magnetic nanostructures},}\ }\href
	{https://doi.org/10.1103/PhysRevLett.106.100501} {\bibfield  {journal}
		{\bibinfo  {journal} {Phys. Rev. Lett.}\ }\textbf {\bibinfo {volume} {106}},\
		\bibinfo {pages} {100501} (\bibinfo {year} {2011})}\BibitemShut {NoStop}%
	\bibitem [{\citenamefont {Gauger}\ \emph {et~al.}(2011)\citenamefont {Gauger},
		\citenamefont {Rieper}, \citenamefont {Morton}, \citenamefont {Benjamin},\
		and\ \citenamefont {Vedral}}]{GaugerErikM2011PRL}%
	\BibitemOpen
	\bibfield  {author} {\bibinfo {author} {\bibfnamefont {E.~M.}\ \bibnamefont
			{Gauger}}, \bibinfo {author} {\bibfnamefont {E.}~\bibnamefont {Rieper}},
		\bibinfo {author} {\bibfnamefont {J.~J.~L.}\ \bibnamefont {Morton}}, \bibinfo
		{author} {\bibfnamefont {S.~C.}\ \bibnamefont {Benjamin}},\ and\ \bibinfo
		{author} {\bibfnamefont {V.}~\bibnamefont {Vedral}},\ }\bibfield  {title}
	{\enquote {\bibinfo {title} {Sustained quantum coherence and entanglement in
				the avian compass},}\ }\href {https://doi.org/10.1103/PhysRevLett.106.040503}
	{\bibfield  {journal} {\bibinfo  {journal} {Phys. Rev. Lett.}\ }\textbf
		{\bibinfo {volume} {106}},\ \bibinfo {pages} {040503} (\bibinfo {year}
		{2011})}\BibitemShut {NoStop}%
	\bibitem [{\citenamefont {Yang}, \citenamefont {Ai},\ and\ \citenamefont
		{Sun}(2012)}]{YangLiPing2012PRA}%
	\BibitemOpen
	\bibfield  {author} {\bibinfo {author} {\bibfnamefont {L.-P.}\ \bibnamefont
			{Yang}}, \bibinfo {author} {\bibfnamefont {Q.}~\bibnamefont {Ai}},\ and\
		\bibinfo {author} {\bibfnamefont {C.~P.}\ \bibnamefont {Sun}},\ }\bibfield
	{title} {\enquote {\bibinfo {title} {Generalized {H}olstein model for
				spin-dependent electron-transfer reactions},}\ }\href
	{https://doi.org/10.1103/PhysRevA.85.032707} {\bibfield  {journal} {\bibinfo
			{journal} {Phys. Rev. A}\ }\textbf {\bibinfo {volume} {85}},\ \bibinfo
		{pages} {032707} (\bibinfo {year} {2012})}\BibitemShut {NoStop}%
	\bibitem [{\citenamefont {Zhang}\ \emph {et~al.}(2021)\citenamefont {Zhang},
		\citenamefont {Tao}, \citenamefont {He}, \citenamefont {Chen}, \citenamefont
		{Kong}, \citenamefont {Deng}, \citenamefont {Lambert},\ and\ \citenamefont
		{Ai}}]{ZhangNaNa2021FOP}%
	\BibitemOpen
	\bibfield  {author} {\bibinfo {author} {\bibfnamefont {N.-N.}\ \bibnamefont
			{Zhang}}, \bibinfo {author} {\bibfnamefont {M.-J.}\ \bibnamefont {Tao}},
		\bibinfo {author} {\bibfnamefont {W.-T.}\ \bibnamefont {He}}, \bibinfo
		{author} {\bibfnamefont {X.-Y.}\ \bibnamefont {Chen}}, \bibinfo {author}
		{\bibfnamefont {X.-Y.}\ \bibnamefont {Kong}}, \bibinfo {author}
		{\bibfnamefont {F.-G.}\ \bibnamefont {Deng}}, \bibinfo {author}
		{\bibfnamefont {N.}~\bibnamefont {Lambert}},\ and\ \bibinfo {author}
		{\bibfnamefont {Q.}~\bibnamefont {Ai}},\ }\bibfield  {title} {\enquote
		{\bibinfo {title} {Efficient quantum simulation of open quantum dynamics at
				various {H}amiltonians and spectral densities},}\ }\href
	{https://doi.org/10.1007/s11467-021-1064-y} {\bibfield  {journal} {\bibinfo
			{journal} {Front. Phys.}\ }\textbf {\bibinfo {volume} {16}},\ \bibinfo
		{pages} {1--14} (\bibinfo {year} {2021})}\BibitemShut {NoStop}%
	\bibitem [{\citenamefont {Wang}\ \emph {et~al.}(2018)\citenamefont {Wang},
		\citenamefont {Tao}, \citenamefont {Ai}, \citenamefont {Xin}, \citenamefont
		{Lambert}, \citenamefont {Ruan}, \citenamefont {Cheng}, \citenamefont {Nori},
		\citenamefont {Deng},\ and\ \citenamefont {Long}}]{WangBiXue2018npj}%
	\BibitemOpen
	\bibfield  {author} {\bibinfo {author} {\bibfnamefont {B.-X.}\ \bibnamefont
			{Wang}}, \bibinfo {author} {\bibfnamefont {M.-J.}\ \bibnamefont {Tao}},
		\bibinfo {author} {\bibfnamefont {Q.}~\bibnamefont {Ai}}, \bibinfo {author}
		{\bibfnamefont {T.}~\bibnamefont {Xin}}, \bibinfo {author} {\bibfnamefont
			{N.}~\bibnamefont {Lambert}}, \bibinfo {author} {\bibfnamefont
			{D.}~\bibnamefont {Ruan}}, \bibinfo {author} {\bibfnamefont {Y.-C.}\
			\bibnamefont {Cheng}}, \bibinfo {author} {\bibfnamefont {F.}~\bibnamefont
			{Nori}}, \bibinfo {author} {\bibfnamefont {F.-G.}\ \bibnamefont {Deng}},\
		and\ \bibinfo {author} {\bibfnamefont {G.-L.}\ \bibnamefont {Long}},\
	}\bibfield  {title} {\enquote {\bibinfo {title} {Efficient quantum simulation
				of photosynthetic light harvesting},}\ }\href
	{https://doi.org/10.1038/s41534-018-0102-2} {\bibfield  {journal} {\bibinfo
			{journal} {npj Quantum Inf.}\ }\textbf {\bibinfo {volume} {4}},\ \bibinfo
		{pages} {52} (\bibinfo {year} {2018})}\BibitemShut {NoStop}%
	\bibitem [{\citenamefont {Chen}\ \emph {et~al.}(2022)\citenamefont {Chen},
		\citenamefont {Zhang}, \citenamefont {He}, \citenamefont {Kong},
		\citenamefont {Tao}, \citenamefont {Deng}, \citenamefont {Ai},\ and\
		\citenamefont {Long}}]{ChenXinYu2022npj}%
	\BibitemOpen
	\bibfield  {author} {\bibinfo {author} {\bibfnamefont {X.-Y.}\ \bibnamefont
			{Chen}}, \bibinfo {author} {\bibfnamefont {N.-N.}\ \bibnamefont {Zhang}},
		\bibinfo {author} {\bibfnamefont {W.-T.}\ \bibnamefont {He}}, \bibinfo
		{author} {\bibfnamefont {X.-Y.}\ \bibnamefont {Kong}}, \bibinfo {author}
		{\bibfnamefont {M.-J.}\ \bibnamefont {Tao}}, \bibinfo {author} {\bibfnamefont
			{F.-G.}\ \bibnamefont {Deng}}, \bibinfo {author} {\bibfnamefont
			{Q.}~\bibnamefont {Ai}},\ and\ \bibinfo {author} {\bibfnamefont {G.-L.}\
			\bibnamefont {Long}},\ }\bibfield  {title} {\enquote {\bibinfo {title}
			{Global correlation and local information flows in controllable
				non-{M}arkovian open quantum dynamics},}\ }\href
	{https://doi.org/10.1038/s41534-022-00537-z} {\bibfield  {journal} {\bibinfo
			{journal} {npj Quantum Inf.}\ }\textbf {\bibinfo {volume} {8}},\ \bibinfo
		{pages} {22} (\bibinfo {year} {2022})}\BibitemShut {NoStop}%
	\bibitem [{\citenamefont {Makri}\ \emph {et~al.}(1996)\citenamefont {Makri},
		\citenamefont {Sim}, \citenamefont {Makarov},\ and\ \citenamefont
		{Topaler}}]{MakriNancy1996PNAS}%
	\BibitemOpen
	\bibfield  {author} {\bibinfo {author} {\bibfnamefont {N.}~\bibnamefont
			{Makri}}, \bibinfo {author} {\bibfnamefont {E.}~\bibnamefont {Sim}}, \bibinfo
		{author} {\bibfnamefont {D.~E.}\ \bibnamefont {Makarov}},\ and\ \bibinfo
		{author} {\bibfnamefont {M.}~\bibnamefont {Topaler}},\ }\bibfield  {title}
	{\enquote {\bibinfo {title} {Long-time quantum simulation of the primary
				charge separation in bacterial photosynthesis},}\ }\href
	{https://doi.org/10.1073/pnas.93.9.3926} {\bibfield  {journal} {\bibinfo
			{journal} {Proc. Natl. Acad. Sci. U.S.A.}\ }\textbf {\bibinfo {volume}
			{93}},\ \bibinfo {pages} {3926--3931} (\bibinfo {year} {1996})}\BibitemShut
	{NoStop}%
	\bibitem [{\citenamefont {Calvin}(1978)}]{CalvinMelvin1978ACR}%
	\BibitemOpen
	\bibfield  {author} {\bibinfo {author} {\bibfnamefont {M.}~\bibnamefont
			{Calvin}},\ }\bibfield  {title} {\enquote {\bibinfo {title} {Simulating
				photosynthetic quantum conversion},}\ }\href
	{https://doi.org/10.1021/ar50130a001} {\bibfield  {journal} {\bibinfo
			{journal} {Acc. Chem. Res.}\ }\textbf {\bibinfo {volume} {11}},\ \bibinfo
		{pages} {369--374} (\bibinfo {year} {1978})}\BibitemShut {NoStop}%
	\bibitem [{\citenamefont {Liguori}\ \emph {et~al.}(2020)\citenamefont
		{Liguori}, \citenamefont {Croce}, \citenamefont {Marrink},\ and\
		\citenamefont {Thallmair}}]{LiguoriNicoletta2020PR}%
	\BibitemOpen
	\bibfield  {author} {\bibinfo {author} {\bibfnamefont {N.}~\bibnamefont
			{Liguori}}, \bibinfo {author} {\bibfnamefont {R.}~\bibnamefont {Croce}},
		\bibinfo {author} {\bibfnamefont {S.~J.}\ \bibnamefont {Marrink}},\ and\
		\bibinfo {author} {\bibfnamefont {S.}~\bibnamefont {Thallmair}},\ }\bibfield
	{title} {\enquote {\bibinfo {title} {Molecular dynamics simulations in
				photosynthesis},}\ }\href
	{https://doi.org/https://doi.org/10.1007/s11120-020-00741-y} {\bibfield
		{journal} {\bibinfo  {journal} {Photosynth. Res.}\ }\textbf {\bibinfo
			{volume} {144}},\ \bibinfo {pages} {273--295} (\bibinfo {year}
		{2020})}\BibitemShut {NoStop}%
	\bibitem [{\citenamefont {Gorman}\ \emph {et~al.}(2018)\citenamefont {Gorman},
		\citenamefont {Hemmerling}, \citenamefont {Megidish}, \citenamefont
		{Moeller}, \citenamefont {Schindler}, \citenamefont {Sarovar},\ and\
		\citenamefont {Haeffner}}]{Gorman2018PRX}%
	\BibitemOpen
	\bibfield  {author} {\bibinfo {author} {\bibfnamefont {D.~J.}\ \bibnamefont
			{Gorman}}, \bibinfo {author} {\bibfnamefont {B.}~\bibnamefont {Hemmerling}},
		\bibinfo {author} {\bibfnamefont {E.}~\bibnamefont {Megidish}}, \bibinfo
		{author} {\bibfnamefont {S.~A.}\ \bibnamefont {Moeller}}, \bibinfo {author}
		{\bibfnamefont {P.}~\bibnamefont {Schindler}}, \bibinfo {author}
		{\bibfnamefont {M.}~\bibnamefont {Sarovar}},\ and\ \bibinfo {author}
		{\bibfnamefont {H.}~\bibnamefont {Haeffner}},\ }\bibfield  {title} {\enquote
		{\bibinfo {title} {Engineering vibrationally assisted energy transfer in a
				trapped-ion quantum simulator},}\ }\href
	{https://doi.org/10.1103/PhysRevX.8.011038} {\bibfield  {journal} {\bibinfo
			{journal} {Phys. Rev. X}\ }\textbf {\bibinfo {volume} {8}},\ \bibinfo {pages}
		{011038} (\bibinfo {year} {2018})}\BibitemShut {NoStop}%
	\bibitem [{\citenamefont {Cirac}\ and\ \citenamefont
		{Zoller}(1995)}]{Cirac1995PRL}%
	\BibitemOpen
	\bibfield  {author} {\bibinfo {author} {\bibfnamefont {J.~I.}\ \bibnamefont
			{Cirac}}\ and\ \bibinfo {author} {\bibfnamefont {P.}~\bibnamefont {Zoller}},\
	}\bibfield  {title} {\enquote {\bibinfo {title} {Quantum computations with
				cold trapped ions},}\ }\href {https://doi.org/10.1103/PhysRevLett.74.4091}
	{\bibfield  {journal} {\bibinfo  {journal} {Phys. Rev. Lett.}\ }\textbf
		{\bibinfo {volume} {74}},\ \bibinfo {pages} {4091--4094} (\bibinfo {year}
		{1995})}\BibitemShut {NoStop}%
	\bibitem [{\citenamefont {H{\"a}ffner}, \citenamefont {Roos},\ and\
		\citenamefont {Blatt}(2008)}]{Haffner2008PR}%
	\BibitemOpen
	\bibfield  {author} {\bibinfo {author} {\bibfnamefont {H.}~\bibnamefont
			{H{\"a}ffner}}, \bibinfo {author} {\bibfnamefont {C.~F.}\ \bibnamefont
			{Roos}},\ and\ \bibinfo {author} {\bibfnamefont {R.}~\bibnamefont {Blatt}},\
	}\bibfield  {title} {\enquote {\bibinfo {title} {Quantum computing with
				trapped ions},}\ }\href
	{https://doi.org/https://doi.org/10.1016/j.physrep.2008.09.003} {\bibfield
		{journal} {\bibinfo  {journal} {Phys. Rep.}\ }\textbf {\bibinfo {volume}
			{469}},\ \bibinfo {pages} {155--203} (\bibinfo {year} {2008})}\BibitemShut
	{NoStop}%
	\bibitem [{\citenamefont {Blatt}\ and\ \citenamefont
		{Roos}(2012)}]{Blatt2012NP}%
	\BibitemOpen
	\bibfield  {author} {\bibinfo {author} {\bibfnamefont {R.}~\bibnamefont
			{Blatt}}\ and\ \bibinfo {author} {\bibfnamefont {C.~F.}\ \bibnamefont
			{Roos}},\ }\bibfield  {title} {\enquote {\bibinfo {title} {Quantum
				simulations with trapped ions},}\ }\href
	{https://doi.org/https://doi.org/10.1038/nphys2252} {\bibfield  {journal}
		{\bibinfo  {journal} {Nat. Phys.}\ }\textbf {\bibinfo {volume} {8}},\
		\bibinfo {pages} {277--284} (\bibinfo {year} {2012})}\BibitemShut {NoStop}%
	\bibitem [{\citenamefont {Zhang}\ \emph {et~al.}(2023)\citenamefont {Zhang},
		\citenamefont {Hu}, \citenamefont {Wang},\ and\ \citenamefont
		{Kais}}]{Zhang2023JPCL}%
	\BibitemOpen
	\bibfield  {author} {\bibinfo {author} {\bibfnamefont {Y.}~\bibnamefont
			{Zhang}}, \bibinfo {author} {\bibfnamefont {Z.}~\bibnamefont {Hu}}, \bibinfo
		{author} {\bibfnamefont {Y.}~\bibnamefont {Wang}},\ and\ \bibinfo {author}
		{\bibfnamefont {S.}~\bibnamefont {Kais}},\ }\bibfield  {title} {\enquote
		{\bibinfo {title} {Quantum simulation of the radical pair dynamics of the
				avian compass},}\ }\href {https://doi.org/10.1021/acs.jpclett.2c03617}
	{\bibfield  {journal} {\bibinfo  {journal} {J. Phys. Chem. Lett.}\ }\textbf
		{\bibinfo {volume} {14}},\ \bibinfo {pages} {832--837} (\bibinfo {year}
		{2023})}\BibitemShut {NoStop}%
	\bibitem [{\citenamefont {Muniz}\ \emph {et~al.}(2020)\citenamefont {Muniz},
		\citenamefont {Barberena}, \citenamefont {Lewis-Swan}, \citenamefont {Young},
		\citenamefont {Cline}, \citenamefont {Rey},\ and\ \citenamefont
		{Thompson}}]{Muniz2020Nature}%
	\BibitemOpen
	\bibfield  {author} {\bibinfo {author} {\bibfnamefont {J.~A.}\ \bibnamefont
			{Muniz}}, \bibinfo {author} {\bibfnamefont {D.}~\bibnamefont {Barberena}},
		\bibinfo {author} {\bibfnamefont {R.~J.}\ \bibnamefont {Lewis-Swan}},
		\bibinfo {author} {\bibfnamefont {D.~J.}\ \bibnamefont {Young}}, \bibinfo
		{author} {\bibfnamefont {J.~R.}\ \bibnamefont {Cline}}, \bibinfo {author}
		{\bibfnamefont {A.~M.}\ \bibnamefont {Rey}},\ and\ \bibinfo {author}
		{\bibfnamefont {J.~K.}\ \bibnamefont {Thompson}},\ }\bibfield  {title}
	{\enquote {\bibinfo {title} {Exploring dynamical phase transitions with cold
				atoms in an optical cavity},}\ }\href
	{https://doi.org/https://doi.org/10.1038/s41586-020-2224-x} {\bibfield
		{journal} {\bibinfo  {journal} {Nature}\ }\textbf {\bibinfo {volume} {580}},\
		\bibinfo {pages} {602--607} (\bibinfo {year} {2020})}\BibitemShut {NoStop}%
	\bibitem [{\citenamefont {Cai}\ \emph {et~al.}(2012)\citenamefont {Cai},
		\citenamefont {Ai}, \citenamefont {Quan},\ and\ \citenamefont
		{Sun}}]{CaiCY22012PRA}%
	\BibitemOpen
	\bibfield  {author} {\bibinfo {author} {\bibfnamefont {C.~Y.}\ \bibnamefont
			{Cai}}, \bibinfo {author} {\bibfnamefont {Q.}~\bibnamefont {Ai}}, \bibinfo
		{author} {\bibfnamefont {H.~T.}\ \bibnamefont {Quan}},\ and\ \bibinfo
		{author} {\bibfnamefont {C.~P.}\ \bibnamefont {Sun}},\ }\bibfield  {title}
	{\enquote {\bibinfo {title} {Sensitive chemical compass assisted by quantum
				criticality},}\ }\href {https://doi.org/10.1103/PhysRevA.85.022315}
	{\bibfield  {journal} {\bibinfo  {journal} {Phys. Rev. A}\ }\textbf {\bibinfo
			{volume} {85}},\ \bibinfo {pages} {022315} (\bibinfo {year}
		{2012})}\BibitemShut {NoStop}%
	\bibitem [{\citenamefont {Wu}\ \emph {et~al.}(2022)\citenamefont {Wu},
		\citenamefont {Hu}, \citenamefont {Zhu}, \citenamefont {Deng},\ and\
		\citenamefont {Ai}}]{WuJiaYi2022JPCB}%
	\BibitemOpen
	\bibfield  {author} {\bibinfo {author} {\bibfnamefont {J.-Y.}\ \bibnamefont
			{Wu}}, \bibinfo {author} {\bibfnamefont {X.-Y.}\ \bibnamefont {Hu}}, \bibinfo
		{author} {\bibfnamefont {H.-Y.}\ \bibnamefont {Zhu}}, \bibinfo {author}
		{\bibfnamefont {R.-Q.}\ \bibnamefont {Deng}},\ and\ \bibinfo {author}
		{\bibfnamefont {Q.}~\bibnamefont {Ai}},\ }\bibfield  {title} {\enquote
		{\bibinfo {title} {A bionic compass based on multiradicals},}\ }\href
	{https://doi.org/10.1021/acs.jpcb.2c02711} {\bibfield  {journal} {\bibinfo
			{journal} {J. Phys. Chem. B}\ }\textbf {\bibinfo {volume} {126}},\ \bibinfo
		{pages} {10327--10334} (\bibinfo {year} {2022})}\BibitemShut {NoStop}%
	\bibitem [{\citenamefont {Creatore}\ \emph {et~al.}(2013)\citenamefont
		{Creatore}, \citenamefont {Parker}, \citenamefont {Emmott},\ and\
		\citenamefont {Chin}}]{Creatore2013PRL}%
	\BibitemOpen
	\bibfield  {author} {\bibinfo {author} {\bibfnamefont {C.}~\bibnamefont
			{Creatore}}, \bibinfo {author} {\bibfnamefont {M.~A.}\ \bibnamefont
			{Parker}}, \bibinfo {author} {\bibfnamefont {S.}~\bibnamefont {Emmott}},\
		and\ \bibinfo {author} {\bibfnamefont {A.~W.}\ \bibnamefont {Chin}},\
	}\bibfield  {title} {\enquote {\bibinfo {title} {Efficient biologically
				inspired photocell enhanced by delocalized quantum states},}\ }\href
	{https://doi.org/10.1103/PhysRevLett.111.253601} {\bibfield  {journal}
		{\bibinfo  {journal} {Phys. Rev. Lett.}\ }\textbf {\bibinfo {volume} {111}},\
		\bibinfo {pages} {253601} (\bibinfo {year} {2013})}\BibitemShut {NoStop}%
	\bibitem [{\citenamefont {Xiao}\ \emph {et~al.}(2020)\citenamefont {Xiao},
		\citenamefont {Hu}, \citenamefont {Cai},\ and\ \citenamefont
		{Zhao}}]{Xiao2020PRL}%
	\BibitemOpen
	\bibfield  {author} {\bibinfo {author} {\bibfnamefont {D.-W.}\ \bibnamefont
			{Xiao}}, \bibinfo {author} {\bibfnamefont {W.-H.}\ \bibnamefont {Hu}},
		\bibinfo {author} {\bibfnamefont {Y.}~\bibnamefont {Cai}},\ and\ \bibinfo
		{author} {\bibfnamefont {N.}~\bibnamefont {Zhao}},\ }\bibfield  {title}
	{\enquote {\bibinfo {title} {Magnetic noise enabled biocompass},}\ }\href
	{https://doi.org/10.1103/PhysRevLett.124.128101} {\bibfield  {journal}
		{\bibinfo  {journal} {Phys. Rev. Lett.}\ }\textbf {\bibinfo {volume} {124}},\
		\bibinfo {pages} {128101} (\bibinfo {year} {2020})}\BibitemShut {NoStop}%
	\bibitem [{\citenamefont {Tao}\ \emph {et~al.}(2020{\natexlab{b}})\citenamefont
		{Tao}, \citenamefont {Zhang}, \citenamefont {Wen}, \citenamefont {Deng},
		\citenamefont {Ai},\ and\ \citenamefont {Long}}]{TaoMingJie2020SciBull}%
	\BibitemOpen
	\bibfield  {author} {\bibinfo {author} {\bibfnamefont {M.-J.}\ \bibnamefont
			{Tao}}, \bibinfo {author} {\bibfnamefont {N.-N.}\ \bibnamefont {Zhang}},
		\bibinfo {author} {\bibfnamefont {P.-Y.}\ \bibnamefont {Wen}}, \bibinfo
		{author} {\bibfnamefont {F.-G.}\ \bibnamefont {Deng}}, \bibinfo {author}
		{\bibfnamefont {Q.}~\bibnamefont {Ai}},\ and\ \bibinfo {author}
		{\bibfnamefont {G.-L.}\ \bibnamefont {Long}},\ }\bibfield  {title} {\enquote
		{\bibinfo {title} {Coherent and incoherent theories for photosynthetic energy
				transfer},}\ }\href
	{https://doi.org/https://doi.org/10.1016/j.scib.2019.12.009} {\bibfield
		{journal} {\bibinfo  {journal} {Sci. Bull.}\ }\textbf {\bibinfo {volume}
			{65}},\ \bibinfo {pages} {318--328} (\bibinfo {year}
		{2020}{\natexlab{b}})}\BibitemShut {NoStop}%
	\bibitem [{\citenamefont {Sun}\ \emph {et~al.}(2023)\citenamefont {Sun},
		\citenamefont {Yao}, \citenamefont {Ai},\ and\ \citenamefont
		{Cheng}}]{SunZongHao2023AQT}%
	\BibitemOpen
	\bibfield  {author} {\bibinfo {author} {\bibfnamefont {Z.-H.}\ \bibnamefont
			{Sun}}, \bibinfo {author} {\bibfnamefont {Y.-X.}\ \bibnamefont {Yao}},
		\bibinfo {author} {\bibfnamefont {Q.}~\bibnamefont {Ai}},\ and\ \bibinfo
		{author} {\bibfnamefont {Y.-C.}\ \bibnamefont {Cheng}},\ }\bibfield  {title}
	{\enquote {\bibinfo {title} {Theory of center-line slope in 2{D} electronic
				spectroscopy with static disorder},}\ }\href {https://doi.org/DOI:
		10.1002/qute.202300163} {\bibfield  {journal} {\bibinfo  {journal} {Adv.
				Quantum Technol.}\ }\textbf {\bibinfo {volume} {8}},\ \bibinfo {pages}
		{2300163} (\bibinfo {year} {2023})}\BibitemShut {NoStop}%
	\bibitem [{\citenamefont {Zhang}\ \emph {et~al.}(2025)\citenamefont {Zhang},
		\citenamefont {Yao}, \citenamefont {Huang}, \citenamefont {Jin},\ and\
		\citenamefont {Ai}}]{ZhangHaoYue2025JCTC}%
	\BibitemOpen
	\bibfield  {author} {\bibinfo {author} {\bibfnamefont {H.-Y.}\ \bibnamefont
			{Zhang}}, \bibinfo {author} {\bibfnamefont {Y.-X.}\ \bibnamefont {Yao}},
		\bibinfo {author} {\bibfnamefont {B.-Y.}\ \bibnamefont {Huang}}, \bibinfo
		{author} {\bibfnamefont {J.-Y.-R.}\ \bibnamefont {Jin}},\ and\ \bibinfo
		{author} {\bibfnamefont {Q.}~\bibnamefont {Ai}},\ }\bibfield  {title}
	{\enquote {\bibinfo {title} {Non-{H}ermitian {H}amiltonian approach for
				two-dimensional coherent spectra of driven systems},}\ }\href
	{https://doi.org/10.1021/acs.jctc.4c01737} {\bibfield  {journal} {\bibinfo
			{journal} {J. Chem. Theory Comput.}\ }\textbf {\bibinfo {volume} {21}},\
		\bibinfo {pages} {4067--4079} (\bibinfo {year} {2025})}\BibitemShut {NoStop}%
	\bibitem [{\citenamefont {Jin}\ \emph {et~al.}(2025)\citenamefont {Jin},
		\citenamefont {Zhang}, \citenamefont {Yao}, \citenamefont {Chen},\ and\
		\citenamefont {Ai}}]{JinJingYiRan2025JCP}%
	\BibitemOpen
	\bibfield  {author} {\bibinfo {author} {\bibfnamefont {J.-Y.-R.}\
			\bibnamefont {Jin}}, \bibinfo {author} {\bibfnamefont {H.-Y.}\ \bibnamefont
			{Zhang}}, \bibinfo {author} {\bibfnamefont {Y.-X.}\ \bibnamefont {Yao}},
		\bibinfo {author} {\bibfnamefont {R.-H.}\ \bibnamefont {Chen}},\ and\
		\bibinfo {author} {\bibfnamefont {Q.}~\bibnamefont {Ai}},\ }\bibfield
	{title} {\enquote {\bibinfo {title} {Observing quantum coherent oscillations
				in a three-level atom via electromagnetically induced transparency by
				two-dimensional spectroscopy},}\ }\href {https://doi.org/10.1063/5.0238336}
	{\bibfield  {journal} {\bibinfo  {journal} {J. Chem. Phys.}\ }\textbf
		{\bibinfo {volume} {162}},\ \bibinfo {pages} {014112} (\bibinfo {year}
		{2025})}\BibitemShut {NoStop}%
	\bibitem [{\citenamefont {Weng}(2018)}]{Weng2018CJCP}%
	\BibitemOpen
	\bibfield  {author} {\bibinfo {author} {\bibfnamefont {Y.-x.}\ \bibnamefont
			{Weng}},\ }\bibfield  {title} {\enquote {\bibinfo {title} {Detection of
				electronic coherence via two-dimensional electronic spectroscopy in condensed
				phase},}\ }\href
	{https://doi.org/https://doi.org/10.1063/1674-0068/31/cjcp1803055} {\bibfield
		{journal} {\bibinfo  {journal} {Chin. J. Chem. Phys.}\ }\textbf {\bibinfo
			{volume} {31}},\ \bibinfo {pages} {135--151} (\bibinfo {year}
		{2018})}\BibitemShut {NoStop}%
	\bibitem [{\citenamefont {Yue}\ \emph {et~al.}(2017)\citenamefont {Yue},
		\citenamefont {Wang}, \citenamefont {Leng}, \citenamefont {Zhu},
		\citenamefont {Chen},\ and\ \citenamefont {Weng}}]{Yue2017CPL}%
	\BibitemOpen
	\bibfield  {author} {\bibinfo {author} {\bibfnamefont {S.}~\bibnamefont
			{Yue}}, \bibinfo {author} {\bibfnamefont {Z.}~\bibnamefont {Wang}}, \bibinfo
		{author} {\bibfnamefont {X.}~\bibnamefont {Leng}}, \bibinfo {author}
		{\bibfnamefont {R.-D.}\ \bibnamefont {Zhu}}, \bibinfo {author} {\bibfnamefont
			{H.-L.}\ \bibnamefont {Chen}},\ and\ \bibinfo {author} {\bibfnamefont
			{Y.-X.}\ \bibnamefont {Weng}},\ }\bibfield  {title} {\enquote {\bibinfo
			{title} {Coupling of multi-vibrational modes in bacteriochlorophyll a in
				solution observed with 2{D} electronic spectroscopy},}\ }\href
	{https://doi.org/https://doi.org/10.1016/j.cplett.2017.03.029} {\bibfield
		{journal} {\bibinfo  {journal} {Chem. Phys. Lett.}\ }\textbf {\bibinfo
			{volume} {683}},\ \bibinfo {pages} {591--597} (\bibinfo {year}
		{2017})}\BibitemShut {NoStop}%
	\bibitem [{\citenamefont {Tiwari}\ \emph {et~al.}(2018)\citenamefont {Tiwari},
		\citenamefont {Matutes}, \citenamefont {Gardiner}, \citenamefont {Jansen},
		\citenamefont {Cogdell},\ and\ \citenamefont {Ogilvie}}]{TiwariVivek2018NC}%
	\BibitemOpen
	\bibfield  {author} {\bibinfo {author} {\bibfnamefont {V.}~\bibnamefont
			{Tiwari}}, \bibinfo {author} {\bibfnamefont {Y.~A.}\ \bibnamefont {Matutes}},
		\bibinfo {author} {\bibfnamefont {A.~T.}\ \bibnamefont {Gardiner}}, \bibinfo
		{author} {\bibfnamefont {T.~L.}\ \bibnamefont {Jansen}}, \bibinfo {author}
		{\bibfnamefont {R.~J.}\ \bibnamefont {Cogdell}},\ and\ \bibinfo {author}
		{\bibfnamefont {J.~P.}\ \bibnamefont {Ogilvie}},\ }\bibfield  {title}
	{\enquote {\bibinfo {title} {Spatially-resolved fluorescence-detected
				two-dimensional electronic spectroscopy probes varying excitonic structure in
				photosynthetic bacteria},}\ }\href
	{https://doi.org/https://doi.org/10.1038/s41467-018-06619-x} {\bibfield
		{journal} {\bibinfo  {journal} {Nat. Commun.}\ }\textbf {\bibinfo {volume}
			{9}},\ \bibinfo {pages} {4219} (\bibinfo {year} {2018})}\BibitemShut
	{NoStop}%
	\bibitem [{\citenamefont {Ginsberg}, \citenamefont {Cheng},\ and\ \citenamefont
		{Fleming}(2009)}]{GinsbergNaomiS2009ACR}%
	\BibitemOpen
	\bibfield  {author} {\bibinfo {author} {\bibfnamefont {N.~S.}\ \bibnamefont
			{Ginsberg}}, \bibinfo {author} {\bibfnamefont {Y.-C.}\ \bibnamefont
			{Cheng}},\ and\ \bibinfo {author} {\bibfnamefont {G.~R.}\ \bibnamefont
			{Fleming}},\ }\bibfield  {title} {\enquote {\bibinfo {title} {Two-dimensional
				electronic spectroscopy of molecular aggregates},}\ }\href
	{https://doi.org/10.1021/ar9001075} {\bibfield  {journal} {\bibinfo
			{journal} {Acc. Chem. Res.}\ }\textbf {\bibinfo {volume} {42}},\ \bibinfo
		{pages} {1352--1363} (\bibinfo {year} {2009})}\BibitemShut {NoStop}%
	\bibitem [{\citenamefont {Schlau-Cohen}, \citenamefont {Ishizaki},\ and\
		\citenamefont {Fleming}(2011)}]{GabrielaSSchlauCohen2011CP}%
	\BibitemOpen
	\bibfield  {author} {\bibinfo {author} {\bibfnamefont {G.~S.}\ \bibnamefont
			{Schlau-Cohen}}, \bibinfo {author} {\bibfnamefont {A.}~\bibnamefont
			{Ishizaki}},\ and\ \bibinfo {author} {\bibfnamefont {G.~R.}\ \bibnamefont
			{Fleming}},\ }\bibfield  {title} {\enquote {\bibinfo {title} {Two-dimensional
				electronic spectroscopy and photosynthesis: {F}undamentals and applications
				to photosynthetic light-harvesting},}\ }\href
	{https://doi.org/https://doi.org/10.1016/j.chemphys.2011.04.025} {\bibfield
		{journal} {\bibinfo  {journal} {Chem. Phys.}\ }\textbf {\bibinfo {volume}
			{386}},\ \bibinfo {pages} {1--22} (\bibinfo {year} {2011})}\BibitemShut
	{NoStop}%
	\bibitem [{\citenamefont {Lewis}\ and\ \citenamefont
		{Ogilvie}(2012)}]{LewisKristinLM2012JPCL}%
	\BibitemOpen
	\bibfield  {author} {\bibinfo {author} {\bibfnamefont {K.~L.~M.}\
			\bibnamefont {Lewis}}\ and\ \bibinfo {author} {\bibfnamefont {J.~P.}\
			\bibnamefont {Ogilvie}},\ }\bibfield  {title} {\enquote {\bibinfo {title}
			{Probing photosynthetic energy and charge transfer with two-dimensional
				electronic spectroscopy},}\ }\href {https://doi.org/10.1021/jz201592v}
	{\bibfield  {journal} {\bibinfo  {journal} {J. Phys. Chem. Lett.}\ }\textbf
		{\bibinfo {volume} {3}},\ \bibinfo {pages} {503--510} (\bibinfo {year}
		{2012})}\BibitemShut {NoStop}%
	\bibitem [{\citenamefont {Zigmantas}\ \emph {et~al.}(2006)\citenamefont
		{Zigmantas}, \citenamefont {Read}, \citenamefont {Man{\v{c}}al},
		\citenamefont {Brixner}, \citenamefont {Gardiner}, \citenamefont {Cogdell},\
		and\ \citenamefont {Fleming}}]{Zigmantas2006PANS}%
	\BibitemOpen
	\bibfield  {author} {\bibinfo {author} {\bibfnamefont {D.}~\bibnamefont
			{Zigmantas}}, \bibinfo {author} {\bibfnamefont {E.~L.}\ \bibnamefont {Read}},
		\bibinfo {author} {\bibfnamefont {T.}~\bibnamefont {Man{\v{c}}al}}, \bibinfo
		{author} {\bibfnamefont {T.}~\bibnamefont {Brixner}}, \bibinfo {author}
		{\bibfnamefont {A.~T.}\ \bibnamefont {Gardiner}}, \bibinfo {author}
		{\bibfnamefont {R.~J.}\ \bibnamefont {Cogdell}},\ and\ \bibinfo {author}
		{\bibfnamefont {G.~R.}\ \bibnamefont {Fleming}},\ }\bibfield  {title}
	{\enquote {\bibinfo {title} {Two-dimensional electronic spectroscopy of the
				{B}800--{B}820 light-harvesting complex},}\ }\href
	{https://doi.org/https://doi.org/10.1073/pnas.060296110} {\bibfield
		{journal} {\bibinfo  {journal} {Proc. Natl. Acad. Sci. U.S.A.}\ }\textbf
		{\bibinfo {volume} {103}},\ \bibinfo {pages} {12672--12677} (\bibinfo {year}
		{2006})}\BibitemShut {NoStop}%
	\bibitem [{\citenamefont {Ai}\ \emph {et~al.}(2014)\citenamefont {Ai},
		\citenamefont {Fan}, \citenamefont {Jin},\ and\ \citenamefont
		{Cheng}}]{AiQing2014NJP}%
	\BibitemOpen
	\bibfield  {author} {\bibinfo {author} {\bibfnamefont {Q.}~\bibnamefont
			{Ai}}, \bibinfo {author} {\bibfnamefont {Y.-J.}\ \bibnamefont {Fan}},
		\bibinfo {author} {\bibfnamefont {B.-Y.}\ \bibnamefont {Jin}},\ and\ \bibinfo
		{author} {\bibfnamefont {Y.-C.}\ \bibnamefont {Cheng}},\ }\bibfield  {title}
	{\enquote {\bibinfo {title} {An efficient quantum jump method for coherent
				energy transfer dynamics in photosynthetic systems under the influence of
				laser fields},}\ }\href {https://doi.org/10.1088/1367-2630/16/5/053033}
	{\bibfield  {journal} {\bibinfo  {journal} {New J. Phys.}\ }\textbf {\bibinfo
			{volume} {16}},\ \bibinfo {pages} {053033} (\bibinfo {year}
		{2014})}\BibitemShut {NoStop}%
	\bibitem [{\citenamefont {Ai}\ \emph {et~al.}(2013)\citenamefont {Ai},
		\citenamefont {Yen}, \citenamefont {Jin},\ and\ \citenamefont
		{Cheng}}]{AiQing2013PCL}%
	\BibitemOpen
	\bibfield  {author} {\bibinfo {author} {\bibfnamefont {Q.}~\bibnamefont
			{Ai}}, \bibinfo {author} {\bibfnamefont {T.-C.}\ \bibnamefont {Yen}},
		\bibinfo {author} {\bibfnamefont {B.-Y.}\ \bibnamefont {Jin}},\ and\ \bibinfo
		{author} {\bibfnamefont {Y.-C.}\ \bibnamefont {Cheng}},\ }\bibfield  {title}
	{\enquote {\bibinfo {title} {Clustered geometries exploiting quantum
				coherence effects for efficient energy transfer in light harvesting},}\
	}\href {https://doi.org/10.1021/jz4011477} {\bibfield  {journal} {\bibinfo
			{journal} {J. Phys. Chem. Lett.}\ }\textbf {\bibinfo {volume} {4}},\ \bibinfo
		{pages} {2577--2584} (\bibinfo {year} {2013})}\BibitemShut {NoStop}%
	\bibitem [{\citenamefont {Lee}, \citenamefont {Cheng},\ and\ \citenamefont
		{Fleming}(2007)}]{LeeHohjai2007Science}%
	\BibitemOpen
	\bibfield  {author} {\bibinfo {author} {\bibfnamefont {H.}~\bibnamefont
			{Lee}}, \bibinfo {author} {\bibfnamefont {Y.-C.}\ \bibnamefont {Cheng}},\
		and\ \bibinfo {author} {\bibfnamefont {G.~R.}\ \bibnamefont {Fleming}},\
	}\bibfield  {title} {\enquote {\bibinfo {title} {Coherence dynamics in
				photosynthesis: {P}rotein protection of excitonic coherence},}\ }\href
	{https://www.science.org/doi/abs/10.1126/science.1142188} {\bibfield
		{journal} {\bibinfo  {journal} {Science}\ }\textbf {\bibinfo {volume}
			{316}},\ \bibinfo {pages} {1462--1465} (\bibinfo {year} {2007})}\BibitemShut
	{NoStop}%
	\bibitem [{\citenamefont {Engel}\ \emph {et~al.}(2007)\citenamefont {Engel},
		\citenamefont {Calhoun}, \citenamefont {Read}, \citenamefont {Ahn},
		\citenamefont {Man{\v{c}}al}, \citenamefont {Cheng}, \citenamefont
		{Blankenship},\ and\ \citenamefont {Fleming}}]{EngelGregoryS2007Nature}%
	\BibitemOpen
	\bibfield  {author} {\bibinfo {author} {\bibfnamefont {G.~S.}\ \bibnamefont
			{Engel}}, \bibinfo {author} {\bibfnamefont {T.~R.}\ \bibnamefont {Calhoun}},
		\bibinfo {author} {\bibfnamefont {E.~L.}\ \bibnamefont {Read}}, \bibinfo
		{author} {\bibfnamefont {T.-K.}\ \bibnamefont {Ahn}}, \bibinfo {author}
		{\bibfnamefont {T.}~\bibnamefont {Man{\v{c}}al}}, \bibinfo {author}
		{\bibfnamefont {Y.-C.}\ \bibnamefont {Cheng}}, \bibinfo {author}
		{\bibfnamefont {R.~E.}\ \bibnamefont {Blankenship}},\ and\ \bibinfo {author}
		{\bibfnamefont {G.~R.}\ \bibnamefont {Fleming}},\ }\bibfield  {title}
	{\enquote {\bibinfo {title} {Evidence for wavelike energy transfer through
				quantum coherence in photosynthetic systems},}\ }\href
	{https://doi.org/https://doi.org/10.1038/nature05678} {\bibfield  {journal}
		{\bibinfo  {journal} {Nature}\ }\textbf {\bibinfo {volume} {446}},\ \bibinfo
		{pages} {782--786} (\bibinfo {year} {2007})}\BibitemShut {NoStop}%
	\bibitem [{\citenamefont {Lloyd}(2011)}]{LloydSeth2011JPCS}%
	\BibitemOpen
	\bibfield  {author} {\bibinfo {author} {\bibfnamefont {S.}~\bibnamefont
			{Lloyd}},\ }\bibfield  {title} {\enquote {\bibinfo {title} {Quantum coherence
				in biological systems},}\ }\href
	{https://doi.org/10.1088/1742-6596/302/1/012037} {\bibfield  {journal}
		{\bibinfo  {journal} {J. Phys.: Conf. Ser.}\ }\textbf {\bibinfo {volume}
			{302}},\ \bibinfo {pages} {012037} (\bibinfo {year} {2011})}\BibitemShut
	{NoStop}%
	\bibitem [{\citenamefont {Ishizaki}\ and\ \citenamefont
		{Fleming}(2012)}]{IshizakiAkihito2012ARCMP}%
	\BibitemOpen
	\bibfield  {author} {\bibinfo {author} {\bibfnamefont {A.}~\bibnamefont
			{Ishizaki}}\ and\ \bibinfo {author} {\bibfnamefont {G.~R.}\ \bibnamefont
			{Fleming}},\ }\bibfield  {title} {\enquote {\bibinfo {title} {Quantum
				coherence in photosynthetic light harvesting},}\ }\href
	{https://doi.org/10.1146/annurev-conmatphys-020911-125126} {\bibfield
		{journal} {\bibinfo  {journal} {Annu. Rev. Condens. Matter Phys.}\ }\textbf
		{\bibinfo {volume} {3}},\ \bibinfo {pages} {333--361} (\bibinfo {year}
		{2012})}\BibitemShut {NoStop}%
	\bibitem [{\citenamefont {Kassal}, \citenamefont {Yuen-Zhou},\ and\
		\citenamefont {Rahimi-Keshari}(2013)}]{KassalIvan2013PCL}%
	\BibitemOpen
	\bibfield  {author} {\bibinfo {author} {\bibfnamefont {I.}~\bibnamefont
			{Kassal}}, \bibinfo {author} {\bibfnamefont {J.}~\bibnamefont {Yuen-Zhou}},\
		and\ \bibinfo {author} {\bibfnamefont {S.}~\bibnamefont {Rahimi-Keshari}},\
	}\bibfield  {title} {\enquote {\bibinfo {title} {Does coherence enhance
				transport in photosynthesis?}}\ }\href
	{https://doi.org/dx.doi.org/10.1021/jz301872b} {\bibfield  {journal}
		{\bibinfo  {journal} {J. Phys. Chem. Lett.}\ }\textbf {\bibinfo {volume}
			{4}},\ \bibinfo {pages} {362--367} (\bibinfo {year} {2013})}\BibitemShut
	{NoStop}%
	\bibitem [{\citenamefont {Fleming}, \citenamefont {Scholes},\ and\
		\citenamefont {Cheng}(2011)}]{GrahamRFleming2011PC}%
	\BibitemOpen
	\bibfield  {author} {\bibinfo {author} {\bibfnamefont {G.~R.}\ \bibnamefont
			{Fleming}}, \bibinfo {author} {\bibfnamefont {G.~D.}\ \bibnamefont
			{Scholes}},\ and\ \bibinfo {author} {\bibfnamefont {Y.-C.}\ \bibnamefont
			{Cheng}},\ }\bibfield  {title} {\enquote {\bibinfo {title} {Quantum effects
				in biology},}\ }\href
	{https://doi.org/https://doi.org/10.1016/j.proche.2011.08.011} {\bibfield
		{journal} {\bibinfo  {journal} {Procedia Chem.}\ }\textbf {\bibinfo {volume}
			{3}},\ \bibinfo {pages} {38--57} (\bibinfo {year} {2011})},\ \bibinfo {note}
	{22nd Solvay Conference on Chemistry}\BibitemShut {NoStop}%
	\bibitem [{\citenamefont {Chenu}\ and\ \citenamefont
		{Scholes}(2015)}]{ChenuAurelia2015ARPC}%
	\BibitemOpen
	\bibfield  {author} {\bibinfo {author} {\bibfnamefont {A.}~\bibnamefont
			{Chenu}}\ and\ \bibinfo {author} {\bibfnamefont {G.~D.}\ \bibnamefont
			{Scholes}},\ }\bibfield  {title} {\enquote {\bibinfo {title} {Coherence in
				energy transfer and photosynthesis},}\ }\href
	{https://doi.org/10.1146/annurev-physchem-040214-121713} {\bibfield
		{journal} {\bibinfo  {journal} {Annu. Rev. Phys. Chem.}\ }\textbf {\bibinfo
			{volume} {66}},\ \bibinfo {pages} {69--96} (\bibinfo {year}
		{2015})}\BibitemShut {NoStop}%
	\bibitem [{\citenamefont {Panitchayangkoon}\ \emph {et~al.}(2011)\citenamefont
		{Panitchayangkoon}, \citenamefont {Voronine}, \citenamefont {Abramavicius},
		\citenamefont {Caram}, \citenamefont {Lewis}, \citenamefont {Mukamel},\ and\
		\citenamefont {Engel}}]{PanitchayangkoonGitt2011PNAS}%
	\BibitemOpen
	\bibfield  {author} {\bibinfo {author} {\bibfnamefont {G.}~\bibnamefont
			{Panitchayangkoon}}, \bibinfo {author} {\bibfnamefont {D.~V.}\ \bibnamefont
			{Voronine}}, \bibinfo {author} {\bibfnamefont {D.}~\bibnamefont
			{Abramavicius}}, \bibinfo {author} {\bibfnamefont {J.~R.}\ \bibnamefont
			{Caram}}, \bibinfo {author} {\bibfnamefont {N.~H.}\ \bibnamefont {Lewis}},
		\bibinfo {author} {\bibfnamefont {S.}~\bibnamefont {Mukamel}},\ and\ \bibinfo
		{author} {\bibfnamefont {G.~S.}\ \bibnamefont {Engel}},\ }\bibfield  {title}
	{\enquote {\bibinfo {title} {Direct evidence of quantum transport in
				photosynthetic light-harvesting complexes},}\ }\href
	{https://doi.org/https://doi.org/10.1073/pnas.110523410} {\bibfield
		{journal} {\bibinfo  {journal} {Proc. Natl. Acad. Sci. U.S.A.}\ }\textbf
		{\bibinfo {volume} {108}},\ \bibinfo {pages} {20908--20912} (\bibinfo {year}
		{2011})}\BibitemShut {NoStop}%
	\bibitem [{\citenamefont {Olaya-Castro}\ \emph {et~al.}(2008)\citenamefont
		{Olaya-Castro}, \citenamefont {Lee}, \citenamefont {Olsen},\ and\
		\citenamefont {Johnson}}]{OlayaCastroAlexandra2008PRB}%
	\BibitemOpen
	\bibfield  {author} {\bibinfo {author} {\bibfnamefont {A.}~\bibnamefont
			{Olaya-Castro}}, \bibinfo {author} {\bibfnamefont {C.~F.}\ \bibnamefont
			{Lee}}, \bibinfo {author} {\bibfnamefont {F.~F.}\ \bibnamefont {Olsen}},\
		and\ \bibinfo {author} {\bibfnamefont {N.~F.}\ \bibnamefont {Johnson}},\
	}\bibfield  {title} {\enquote {\bibinfo {title} {Efficiency of energy
				transfer in a light-harvesting system under quantum coherence},}\ }\href
	{https://doi.org/10.1103/PhysRevB.78.085115} {\bibfield  {journal} {\bibinfo
			{journal} {Phys. Rev. B}\ }\textbf {\bibinfo {volume} {78}},\ \bibinfo
		{pages} {085115} (\bibinfo {year} {2008})}\BibitemShut {NoStop}%
	\bibitem [{\citenamefont {Scholes}(2010)}]{ScholesGregoryD2010PCL}%
	\BibitemOpen
	\bibfield  {author} {\bibinfo {author} {\bibfnamefont {G.~D.}\ \bibnamefont
			{Scholes}},\ }\bibfield  {title} {\enquote {\bibinfo {title}
			{Quantum-coherent electronic energy transfer:{D}id nature think of it
				first?}}\ }\href {https://doi.org/0.1021/jz900062f} {\bibfield  {journal}
		{\bibinfo  {journal} {J. Phys. Chem. Lett.}\ }\textbf {\bibinfo {volume}
			{1}},\ \bibinfo {pages} {2--8} (\bibinfo {year} {2010})}\BibitemShut
	{NoStop}%
	\bibitem [{\citenamefont {Zhu}\ \emph {et~al.}(2024)\citenamefont {Zhu},
		\citenamefont {Li}, \citenamefont {Zhen}, \citenamefont {Zou}, \citenamefont
		{Liao}, \citenamefont {Wang}, \citenamefont {Wang}, \citenamefont {Chen},
		\citenamefont {Qin},\ and\ \citenamefont {Weng}}]{Zhu2024NC}%
	\BibitemOpen
	\bibfield  {author} {\bibinfo {author} {\bibfnamefont {R.}~\bibnamefont
			{Zhu}}, \bibinfo {author} {\bibfnamefont {W.}~\bibnamefont {Li}}, \bibinfo
		{author} {\bibfnamefont {Z.}~\bibnamefont {Zhen}}, \bibinfo {author}
		{\bibfnamefont {J.}~\bibnamefont {Zou}}, \bibinfo {author} {\bibfnamefont
			{G.}~\bibnamefont {Liao}}, \bibinfo {author} {\bibfnamefont {J.}~\bibnamefont
			{Wang}}, \bibinfo {author} {\bibfnamefont {Z.}~\bibnamefont {Wang}}, \bibinfo
		{author} {\bibfnamefont {H.}~\bibnamefont {Chen}}, \bibinfo {author}
		{\bibfnamefont {S.}~\bibnamefont {Qin}},\ and\ \bibinfo {author}
		{\bibfnamefont {Y.}~\bibnamefont {Weng}},\ }\bibfield  {title} {\enquote
		{\bibinfo {title} {Quantum phase synchronization via exciton-vibrational
				energy dissipation sustains long-lived coherence in photosynthetic
				antennas},}\ }\href
	{https://doi.org/https://doi.org/10.1038/s41467-024-47560-6} {\bibfield
		{journal} {\bibinfo  {journal} {Nat. Commun.}\ }\textbf {\bibinfo {volume}
			{15}},\ \bibinfo {pages} {3171} (\bibinfo {year} {2024})}\BibitemShut
	{NoStop}%
	\bibitem [{\citenamefont {Abramavicius}, \citenamefont {Valkunas},\ and\
		\citenamefont {van Grondelle}(2004)}]{AbramaviciusDarius2004PCCP}%
	\BibitemOpen
	\bibfield  {author} {\bibinfo {author} {\bibfnamefont {D.}~\bibnamefont
			{Abramavicius}}, \bibinfo {author} {\bibfnamefont {L.}~\bibnamefont
			{Valkunas}},\ and\ \bibinfo {author} {\bibfnamefont {R.}~\bibnamefont {van
				Grondelle}},\ }\bibfield  {title} {\enquote {\bibinfo {title} {Exciton
				dynamics in ring-like photosynthetic light-harvesting complexes: {A} hopping
				model},}\ }\href {https://doi.org/https://doi.org/10.1039/B315252A}
	{\bibfield  {journal} {\bibinfo  {journal} {Phys. Chem. Chem. Phys.}\
		}\textbf {\bibinfo {volume} {6}},\ \bibinfo {pages} {3097--3105} (\bibinfo
		{year} {2004})}\BibitemShut {NoStop}%
	\bibitem [{\citenamefont {Blankenship}(2021)}]{Blankenship2021book}%
	\BibitemOpen
	\bibfield  {author} {\bibinfo {author} {\bibfnamefont {R.~E.}\ \bibnamefont
			{Blankenship}},\ }\href@noop {} {\emph {\bibinfo {title} {Molecular
				mechanisms of photosynthesis}}}\ (\bibinfo  {publisher} {John Wiley \&
		Sons},\ \bibinfo {year} {2021})\BibitemShut {NoStop}%
	\bibitem [{\citenamefont {Blankenship}\ \emph {et~al.}(2011)\citenamefont
		{Blankenship}, \citenamefont {Tiede}, \citenamefont {Barber}, \citenamefont
		{Brudvig}, \citenamefont {Fleming}, \citenamefont {Ghirardi}, \citenamefont
		{Gunner}, \citenamefont {Junge}, \citenamefont {Kramer}, \citenamefont
		{Melis} \emph {et~al.}}]{Blankenship2011Science}%
	\BibitemOpen
	\bibfield  {author} {\bibinfo {author} {\bibfnamefont {R.~E.}\ \bibnamefont
			{Blankenship}}, \bibinfo {author} {\bibfnamefont {D.~M.}\ \bibnamefont
			{Tiede}}, \bibinfo {author} {\bibfnamefont {J.}~\bibnamefont {Barber}},
		\bibinfo {author} {\bibfnamefont {G.~W.}\ \bibnamefont {Brudvig}}, \bibinfo
		{author} {\bibfnamefont {G.}~\bibnamefont {Fleming}}, \bibinfo {author}
		{\bibfnamefont {M.}~\bibnamefont {Ghirardi}}, \bibinfo {author}
		{\bibfnamefont {M.}~\bibnamefont {Gunner}}, \bibinfo {author} {\bibfnamefont
			{W.}~\bibnamefont {Junge}}, \bibinfo {author} {\bibfnamefont {D.~M.}\
			\bibnamefont {Kramer}}, \bibinfo {author} {\bibfnamefont {A.}~\bibnamefont
			{Melis}}, \emph {et~al.},\ }\bibfield  {title} {\enquote {\bibinfo {title}
			{Comparing photosynthetic and photovoltaic efficiencies and recognizing the
				potential for improvement},}\ }\href
	{https://doi.org/10.1126/science.1200165} {\bibfield  {journal} {\bibinfo
			{journal} {Science}\ }\textbf {\bibinfo {volume} {332}},\ \bibinfo {pages}
		{805--809} (\bibinfo {year} {2011})}\BibitemShut {NoStop}%
	\bibitem [{\citenamefont {Romero}, \citenamefont {Novoderezhkin},\ and\
		\citenamefont {Van~Grondelle}(2017)}]{Romero2017Nature}%
	\BibitemOpen
	\bibfield  {author} {\bibinfo {author} {\bibfnamefont {E.}~\bibnamefont
			{Romero}}, \bibinfo {author} {\bibfnamefont {V.~I.}\ \bibnamefont
			{Novoderezhkin}},\ and\ \bibinfo {author} {\bibfnamefont {R.}~\bibnamefont
			{Van~Grondelle}},\ }\bibfield  {title} {\enquote {\bibinfo {title} {Quantum
				design of photosynthesis for bio-inspired solar-energy conversion},}\ }\href
	{https://doi.org/https://doi.org/10.1038/nature22012} {\bibfield  {journal}
		{\bibinfo  {journal} {Nature}\ }\textbf {\bibinfo {volume} {543}},\ \bibinfo
		{pages} {355--365} (\bibinfo {year} {2017})}\BibitemShut {NoStop}%
	\bibitem [{\citenamefont {Gr{\"a}tzel}(1991)}]{Gratzel1991CIC}%
	\BibitemOpen
	\bibfield  {author} {\bibinfo {author} {\bibfnamefont {M.}~\bibnamefont
			{Gr{\"a}tzel}},\ }\bibfield  {title} {\enquote {\bibinfo {title} {The
				artificial leaf, molecular photovoltaics achieve efficient generation of
				electricity from sunlight},}\ }\href
	{https://doi.org/https://doi.org/10.1080/02603599108050599} {\bibfield
		{journal} {\bibinfo  {journal} {Comment. Inorg. Chem.}\ }\textbf {\bibinfo
			{volume} {12}},\ \bibinfo {pages} {93--111} (\bibinfo {year}
		{1991})}\BibitemShut {NoStop}%
	\bibitem [{\citenamefont {Romero}\ \emph {et~al.}(2014)\citenamefont {Romero},
		\citenamefont {Augulis}, \citenamefont {Novoderezhkin}, \citenamefont
		{Ferretti}, \citenamefont {Thieme}, \citenamefont {Zigmantas},\ and\
		\citenamefont {Van~Grondelle}}]{RomeroElisabet2014NP}%
	\BibitemOpen
	\bibfield  {author} {\bibinfo {author} {\bibfnamefont {E.}~\bibnamefont
			{Romero}}, \bibinfo {author} {\bibfnamefont {R.}~\bibnamefont {Augulis}},
		\bibinfo {author} {\bibfnamefont {V.~I.}\ \bibnamefont {Novoderezhkin}},
		\bibinfo {author} {\bibfnamefont {M.}~\bibnamefont {Ferretti}}, \bibinfo
		{author} {\bibfnamefont {J.}~\bibnamefont {Thieme}}, \bibinfo {author}
		{\bibfnamefont {D.}~\bibnamefont {Zigmantas}},\ and\ \bibinfo {author}
		{\bibfnamefont {R.}~\bibnamefont {Van~Grondelle}},\ }\bibfield  {title}
	{\enquote {\bibinfo {title} {Quantum coherence in photosynthesis for
				efficient solar-energy conversion},}\ }\href
	{https://doi.org/https://doi.org/10.1038/nphys3017} {\bibfield  {journal}
		{\bibinfo  {journal} {Nat. Phys.}\ }\textbf {\bibinfo {volume} {10}},\
		\bibinfo {pages} {676--682} (\bibinfo {year} {2014})}\BibitemShut {NoStop}%
	\bibitem [{\citenamefont {Alharbi}\ and\ \citenamefont
		{Kais}(2015)}]{AlharbiFahhadH2015RSER}%
	\BibitemOpen
	\bibfield  {author} {\bibinfo {author} {\bibfnamefont {F.~H.}\ \bibnamefont
			{Alharbi}}\ and\ \bibinfo {author} {\bibfnamefont {S.}~\bibnamefont {Kais}},\
	}\bibfield  {title} {\enquote {\bibinfo {title} {Theoretical limits of
				photovoltaics efficiency and possible improvements by intuitive approaches
				learned from photosynthesis and quantum coherence},}\ }\href
	{https://doi.org/http://dx.doi.org/10.1016/j.rser.2014.11.101} {\bibfield
		{journal} {\bibinfo  {journal} {Renewable Sustainable Energy Rev.}\ }\textbf
		{\bibinfo {volume} {43}},\ \bibinfo {pages} {1073--1089} (\bibinfo {year}
		{2015})}\BibitemShut {NoStop}%
	\bibitem [{\citenamefont {Br{\'e}das}, \citenamefont {Sargent},\ and\
		\citenamefont {Scholes}(2017)}]{BredasJeanLuc2017NM}%
	\BibitemOpen
	\bibfield  {author} {\bibinfo {author} {\bibfnamefont {J.-L.}\ \bibnamefont
			{Br{\'e}das}}, \bibinfo {author} {\bibfnamefont {E.~H.}\ \bibnamefont
			{Sargent}},\ and\ \bibinfo {author} {\bibfnamefont {G.~D.}\ \bibnamefont
			{Scholes}},\ }\bibfield  {title} {\enquote {\bibinfo {title} {Photovoltaic
				concepts inspired by coherence effects in photosynthetic systems},}\ }\href
	{https://doi.org/https://doi.org/10.1038/nmat4767} {\bibfield  {journal}
		{\bibinfo  {journal} {Nat. Mater.}\ }\textbf {\bibinfo {volume} {16}},\
		\bibinfo {pages} {35--44} (\bibinfo {year} {2017})}\BibitemShut {NoStop}%
	\bibitem [{\citenamefont {Yao}\ and\ \citenamefont
		{Ai}(2023)}]{YaoYiXuanADP2023}%
	\BibitemOpen
	\bibfield  {author} {\bibinfo {author} {\bibfnamefont {Y.-X.}\ \bibnamefont
			{Yao}}\ and\ \bibinfo {author} {\bibfnamefont {Q.}~\bibnamefont {Ai}},\
	}\bibfield  {title} {\enquote {\bibinfo {title} {Optical non-reciprocity in
				coupled resonators by detailed balance},}\ }\href
	{https://doi.org/10.1002/andp.202300135} {\bibfield  {journal} {\bibinfo
			{journal} {Ann. Phys.}\ }\textbf {\bibinfo {volume} {535}},\ \bibinfo {pages}
		{2300135} (\bibinfo {year} {2023})}\BibitemShut {NoStop}%
	\bibitem [{\citenamefont {Wiltschko}\ \emph {et~al.}(2010)\citenamefont
		{Wiltschko}, \citenamefont {Stapput}, \citenamefont {Thalau},\ and\
		\citenamefont {Wiltschko}}]{Wiltschko2010JRSI}%
	\BibitemOpen
	\bibfield  {author} {\bibinfo {author} {\bibfnamefont {R.}~\bibnamefont
			{Wiltschko}}, \bibinfo {author} {\bibfnamefont {K.}~\bibnamefont {Stapput}},
		\bibinfo {author} {\bibfnamefont {P.}~\bibnamefont {Thalau}},\ and\ \bibinfo
		{author} {\bibfnamefont {W.}~\bibnamefont {Wiltschko}},\ }\bibfield  {title}
	{\enquote {\bibinfo {title} {Directional orientation of birds by the magnetic
				field under different light conditions},}\ }\href
	{https://doi.org/http://doi.org/10.1098/rsif.2009.0367.focus} {\bibfield
		{journal} {\bibinfo  {journal} {J. R. Soc. Interf.}\ }\textbf {\bibinfo
			{volume} {7}},\ \bibinfo {pages} {163--177} (\bibinfo {year}
		{2010})}\BibitemShut {NoStop}%
	\bibitem [{\citenamefont {Treiber}\ \emph {et~al.}(2012)\citenamefont
		{Treiber}, \citenamefont {Salzer}, \citenamefont {Riegler}, \citenamefont
		{Edelman}, \citenamefont {Sugar}, \citenamefont {Breuss}, \citenamefont
		{Pichler}, \citenamefont {Cadiou}, \citenamefont {Saunders}, \citenamefont
		{Lythgoe} \emph {et~al.}}]{Treiber2012Nature}%
	\BibitemOpen
	\bibfield  {author} {\bibinfo {author} {\bibfnamefont {C.~D.}\ \bibnamefont
			{Treiber}}, \bibinfo {author} {\bibfnamefont {M.~C.}\ \bibnamefont {Salzer}},
		\bibinfo {author} {\bibfnamefont {J.}~\bibnamefont {Riegler}}, \bibinfo
		{author} {\bibfnamefont {N.}~\bibnamefont {Edelman}}, \bibinfo {author}
		{\bibfnamefont {C.}~\bibnamefont {Sugar}}, \bibinfo {author} {\bibfnamefont
			{M.}~\bibnamefont {Breuss}}, \bibinfo {author} {\bibfnamefont
			{P.}~\bibnamefont {Pichler}}, \bibinfo {author} {\bibfnamefont
			{H.}~\bibnamefont {Cadiou}}, \bibinfo {author} {\bibfnamefont
			{M.}~\bibnamefont {Saunders}}, \bibinfo {author} {\bibfnamefont
			{M.}~\bibnamefont {Lythgoe}}, \emph {et~al.},\ }\bibfield  {title} {\enquote
		{\bibinfo {title} {Clusters of iron-rich cells in the upper beak of pigeons
				are macrophages not magnetosensitive neurons},}\ }\href
	{https://doi.org/https://doi.org/10.1038/nature11046} {\bibfield  {journal}
		{\bibinfo  {journal} {Nature}\ }\textbf {\bibinfo {volume} {484}},\ \bibinfo
		{pages} {367--370} (\bibinfo {year} {2012})}\BibitemShut {NoStop}%
	\bibitem [{\citenamefont {Wiltschko}, \citenamefont {Wiltschko},\ and\
		\citenamefont {Munro}(2000)}]{Wiltschko2000Naturwissenschaften}%
	\BibitemOpen
	\bibfield  {author} {\bibinfo {author} {\bibfnamefont {W.}~\bibnamefont
			{Wiltschko}}, \bibinfo {author} {\bibfnamefont {R.}~\bibnamefont
			{Wiltschko}},\ and\ \bibinfo {author} {\bibfnamefont {U.}~\bibnamefont
			{Munro}},\ }\bibfield  {title} {\enquote {\bibinfo {title} {Light-dependent
				magnetoreception in birds: {D}oes directional information change with light
				intensity?}}\ }\href {https://doi.org/https://doi.org/10.1007/s001140050006}
	{\bibfield  {journal} {\bibinfo  {journal} {Naturwissenschaften}\ }\textbf
		{\bibinfo {volume} {87}},\ \bibinfo {pages} {36--40} (\bibinfo {year}
		{2000})}\BibitemShut {NoStop}%
	\bibitem [{\citenamefont {Wiltschko}\ \emph {et~al.}(2002)\citenamefont
		{Wiltschko}, \citenamefont {Traudt}, \citenamefont {G{\"u}nt{\"u}rk{\"u}n},
		\citenamefont {Prior},\ and\ \citenamefont
		{Wiltschko}}]{Wiltschko2002Nature}%
	\BibitemOpen
	\bibfield  {author} {\bibinfo {author} {\bibfnamefont {W.}~\bibnamefont
			{Wiltschko}}, \bibinfo {author} {\bibfnamefont {J.}~\bibnamefont {Traudt}},
		\bibinfo {author} {\bibfnamefont {O.}~\bibnamefont {G{\"u}nt{\"u}rk{\"u}n}},
		\bibinfo {author} {\bibfnamefont {H.}~\bibnamefont {Prior}},\ and\ \bibinfo
		{author} {\bibfnamefont {R.}~\bibnamefont {Wiltschko}},\ }\bibfield  {title}
	{\enquote {\bibinfo {title} {Lateralization of magnetic compass orientation
				in a migratory bird},}\ }\href
	{https://doi.org/https://doi.org/10.1038/nature00958} {\bibfield  {journal}
		{\bibinfo  {journal} {Nature}\ }\textbf {\bibinfo {volume} {419}},\ \bibinfo
		{pages} {467--470} (\bibinfo {year} {2002})}\BibitemShut {NoStop}%
	\bibitem [{\citenamefont {Ritz}\ \emph {et~al.}(2004)\citenamefont {Ritz},
		\citenamefont {Thalau}, \citenamefont {Phillips}, \citenamefont {Wiltschko},\
		and\ \citenamefont {Wiltschko}}]{Ritz2004Nature}%
	\BibitemOpen
	\bibfield  {author} {\bibinfo {author} {\bibfnamefont {T.}~\bibnamefont
			{Ritz}}, \bibinfo {author} {\bibfnamefont {P.}~\bibnamefont {Thalau}},
		\bibinfo {author} {\bibfnamefont {J.~B.}\ \bibnamefont {Phillips}}, \bibinfo
		{author} {\bibfnamefont {R.}~\bibnamefont {Wiltschko}},\ and\ \bibinfo
		{author} {\bibfnamefont {W.}~\bibnamefont {Wiltschko}},\ }\bibfield  {title}
	{\enquote {\bibinfo {title} {Resonance effects indicate a radical-pair
				mechanism for avian magnetic compass},}\ }\href
	{https://doi.org/https://doi.org/10.1038/nature02534} {\bibfield  {journal}
		{\bibinfo  {journal} {Nature}\ }\textbf {\bibinfo {volume} {429}},\ \bibinfo
		{pages} {177--180} (\bibinfo {year} {2004})}\BibitemShut {NoStop}%
	\bibitem [{\citenamefont {Qin}\ \emph {et~al.}(2016)\citenamefont {Qin},
		\citenamefont {Yin}, \citenamefont {Yang}, \citenamefont {Dou}, \citenamefont
		{Liu}, \citenamefont {Zhang}, \citenamefont {Yu}, \citenamefont {Huang},
		\citenamefont {Feng}, \citenamefont {Hao} \emph {et~al.}}]{Qin2016NM}%
	\BibitemOpen
	\bibfield  {author} {\bibinfo {author} {\bibfnamefont {S.}~\bibnamefont
			{Qin}}, \bibinfo {author} {\bibfnamefont {H.}~\bibnamefont {Yin}}, \bibinfo
		{author} {\bibfnamefont {C.}~\bibnamefont {Yang}}, \bibinfo {author}
		{\bibfnamefont {Y.}~\bibnamefont {Dou}}, \bibinfo {author} {\bibfnamefont
			{Z.}~\bibnamefont {Liu}}, \bibinfo {author} {\bibfnamefont {P.}~\bibnamefont
			{Zhang}}, \bibinfo {author} {\bibfnamefont {H.}~\bibnamefont {Yu}}, \bibinfo
		{author} {\bibfnamefont {Y.}~\bibnamefont {Huang}}, \bibinfo {author}
		{\bibfnamefont {J.}~\bibnamefont {Feng}}, \bibinfo {author} {\bibfnamefont
			{J.}~\bibnamefont {Hao}}, \emph {et~al.},\ }\bibfield  {title} {\enquote
		{\bibinfo {title} {A magnetic protein biocompass},}\ }\href
	{https://doi.org/https://doi.org/10.1038/nmat4484} {\bibfield  {journal}
		{\bibinfo  {journal} {Nat. Mater.}\ }\textbf {\bibinfo {volume} {15}},\
		\bibinfo {pages} {217--226} (\bibinfo {year} {2016})}\BibitemShut {NoStop}%
	\bibitem [{\citenamefont {Guo}\ \emph {et~al.}(2021)\citenamefont {Guo},
		\citenamefont {Xu}, \citenamefont {Chen}, \citenamefont {Wang}, \citenamefont
		{Yang}, \citenamefont {Qin}, \citenamefont {Zhao}, \citenamefont {Fei},
		\citenamefont {Zhao}, \citenamefont {Tan} \emph {et~al.}}]{Guo2021SR}%
	\BibitemOpen
	\bibfield  {author} {\bibinfo {author} {\bibfnamefont {Z.}~\bibnamefont
			{Guo}}, \bibinfo {author} {\bibfnamefont {S.}~\bibnamefont {Xu}}, \bibinfo
		{author} {\bibfnamefont {X.}~\bibnamefont {Chen}}, \bibinfo {author}
		{\bibfnamefont {C.}~\bibnamefont {Wang}}, \bibinfo {author} {\bibfnamefont
			{P.}~\bibnamefont {Yang}}, \bibinfo {author} {\bibfnamefont {S.}~\bibnamefont
			{Qin}}, \bibinfo {author} {\bibfnamefont {C.}~\bibnamefont {Zhao}}, \bibinfo
		{author} {\bibfnamefont {F.}~\bibnamefont {Fei}}, \bibinfo {author}
		{\bibfnamefont {X.}~\bibnamefont {Zhao}}, \bibinfo {author} {\bibfnamefont
			{P.-H.}\ \bibnamefont {Tan}}, \emph {et~al.},\ }\bibfield  {title} {\enquote
		{\bibinfo {title} {Modulation of {M}ag{R} magnetic properties via
				iron--sulfur cluster binding},}\ }\href
	{https://doi.org/https://doi.org/10.1038/s41598-021-03344-2} {\bibfield
		{journal} {\bibinfo  {journal} {Sci. Rep.}\ }\textbf {\bibinfo {volume}
			{11}},\ \bibinfo {pages} {23941} (\bibinfo {year} {2021})}\BibitemShut
	{NoStop}%
	\bibitem [{\citenamefont {Zhou}\ \emph {et~al.}(2023)\citenamefont {Zhou},
		\citenamefont {Tong}, \citenamefont {Wei}, \citenamefont {Zhang},
		\citenamefont {Fei}, \citenamefont {Zhou}, \citenamefont {Guo}, \citenamefont
		{Zhang}, \citenamefont {Xu}, \citenamefont {Zhang} \emph
		{et~al.}}]{Zhou2023ZR}%
	\BibitemOpen
	\bibfield  {author} {\bibinfo {author} {\bibfnamefont {Y.}~\bibnamefont
			{Zhou}}, \bibinfo {author} {\bibfnamefont {T.}~\bibnamefont {Tong}}, \bibinfo
		{author} {\bibfnamefont {M.}~\bibnamefont {Wei}}, \bibinfo {author}
		{\bibfnamefont {P.}~\bibnamefont {Zhang}}, \bibinfo {author} {\bibfnamefont
			{F.}~\bibnamefont {Fei}}, \bibinfo {author} {\bibfnamefont {X.}~\bibnamefont
			{Zhou}}, \bibinfo {author} {\bibfnamefont {Z.}~\bibnamefont {Guo}}, \bibinfo
		{author} {\bibfnamefont {J.}~\bibnamefont {Zhang}}, \bibinfo {author}
		{\bibfnamefont {H.}~\bibnamefont {Xu}}, \bibinfo {author} {\bibfnamefont
			{L.}~\bibnamefont {Zhang}}, \emph {et~al.},\ }\bibfield  {title} {\enquote
		{\bibinfo {title} {Towards magnetism in pigeon {M}ag{R}: {I}ron-and
				iron-sulfur binding work indispensably and synergistically},}\ }\href
	{https://doi.org/10.24272/j.issn.2095-8137.2022.423} {\bibfield  {journal}
		{\bibinfo  {journal} {Zoological Research}\ }\textbf {\bibinfo {volume}
			{44}},\ \bibinfo {pages} {142} (\bibinfo {year} {2023})}\BibitemShut
	{NoStop}%
	\bibitem [{\citenamefont {Ritz}, \citenamefont {Adem},\ and\ \citenamefont
		{Schulten}(2000)}]{RitzThorsten2000BJ}%
	\BibitemOpen
	\bibfield  {author} {\bibinfo {author} {\bibfnamefont {T.}~\bibnamefont
			{Ritz}}, \bibinfo {author} {\bibfnamefont {S.}~\bibnamefont {Adem}},\ and\
		\bibinfo {author} {\bibfnamefont {K.}~\bibnamefont {Schulten}},\ }\bibfield
	{title} {\enquote {\bibinfo {title} {A model for photoreceptor-based
				magnetoreception in birds},}\ }\href {https://doi.org/0006-3495/00/02/707/12}
	{\bibfield  {journal} {\bibinfo  {journal} {Biophys. J.}\ }\textbf {\bibinfo
			{volume} {78}},\ \bibinfo {pages} {707--718} (\bibinfo {year}
		{2000})}\BibitemShut {NoStop}%
	\bibitem [{\citenamefont {Bandyopadhyay}, \citenamefont {Paterek},\ and\
		\citenamefont {Kaszlikowski}(2012)}]{Bandyopadhyay2012PRL}%
	\BibitemOpen
	\bibfield  {author} {\bibinfo {author} {\bibfnamefont {J.~N.}\ \bibnamefont
			{Bandyopadhyay}}, \bibinfo {author} {\bibfnamefont {T.}~\bibnamefont
			{Paterek}},\ and\ \bibinfo {author} {\bibfnamefont {D.}~\bibnamefont
			{Kaszlikowski}},\ }\bibfield  {title} {\enquote {\bibinfo {title} {Quantum
				coherence and sensitivity of avian magnetoreception},}\ }\href
	{https://doi.org/10.1103/PhysRevLett.109.110502} {\bibfield  {journal}
		{\bibinfo  {journal} {Phys. Rev. Lett.}\ }\textbf {\bibinfo {volume} {109}},\
		\bibinfo {pages} {110502} (\bibinfo {year} {2012})}\BibitemShut {NoStop}%
	\bibitem [{\citenamefont {Ritz}(2011)}]{RitzThorsten2011PC}%
	\BibitemOpen
	\bibfield  {author} {\bibinfo {author} {\bibfnamefont {T.}~\bibnamefont
			{Ritz}},\ }\bibfield  {title} {\enquote {\bibinfo {title} {Quantum effects in
				biology: {B}ird navigation},}\ }\href
	{https://doi.org/https://doi.org/10.1016/j.proche.2011.08.034} {\bibfield
		{journal} {\bibinfo  {journal} {Procedia Chem.}\ }\textbf {\bibinfo {volume}
			{3}},\ \bibinfo {pages} {262--275} (\bibinfo {year} {2011})}\BibitemShut
	{NoStop}%
	\bibitem [{\citenamefont {Cai}\ and\ \citenamefont
		{Plenio}(2013)}]{CaiJianming2013PRL}%
	\BibitemOpen
	\bibfield  {author} {\bibinfo {author} {\bibfnamefont {J.}~\bibnamefont
			{Cai}}\ and\ \bibinfo {author} {\bibfnamefont {M.~B.}\ \bibnamefont
			{Plenio}},\ }\bibfield  {title} {\enquote {\bibinfo {title} {Chemical compass
				model for avian magnetoreception as a quantum coherent device},}\ }\href
	{https://doi.org/10.1103/PhysRevLett.111.230503} {\bibfield  {journal}
		{\bibinfo  {journal} {Phys. Rev. Lett.}\ }\textbf {\bibinfo {volume} {111}},\
		\bibinfo {pages} {230503} (\bibinfo {year} {2013})}\BibitemShut {NoStop}%
	\bibitem [{\citenamefont {Hiscock}\ \emph {et~al.}(2016)\citenamefont
		{Hiscock}, \citenamefont {Worster}, \citenamefont {Kattnig}, \citenamefont
		{Steers}, \citenamefont {Jin}, \citenamefont {Manolopoulos}, \citenamefont
		{Mouritsen},\ and\ \citenamefont {Hore}}]{Hiscock2016PNAS}%
	\BibitemOpen
	\bibfield  {author} {\bibinfo {author} {\bibfnamefont {H.~G.}\ \bibnamefont
			{Hiscock}}, \bibinfo {author} {\bibfnamefont {S.}~\bibnamefont {Worster}},
		\bibinfo {author} {\bibfnamefont {D.~R.}\ \bibnamefont {Kattnig}}, \bibinfo
		{author} {\bibfnamefont {C.}~\bibnamefont {Steers}}, \bibinfo {author}
		{\bibfnamefont {Y.}~\bibnamefont {Jin}}, \bibinfo {author} {\bibfnamefont
			{D.~E.}\ \bibnamefont {Manolopoulos}}, \bibinfo {author} {\bibfnamefont
			{H.}~\bibnamefont {Mouritsen}},\ and\ \bibinfo {author} {\bibfnamefont
			{P.~J.}\ \bibnamefont {Hore}},\ }\bibfield  {title} {\enquote {\bibinfo
			{title} {The quantum needle of the avian magnetic compass},}\ }\href
	{https://doi.org/https://doi.org/10.1073/pnas.1600341113} {\bibfield
		{journal} {\bibinfo  {journal} {Proc. Natl. Acad. Sci. U.S.A.}\ }\textbf
		{\bibinfo {volume} {113}},\ \bibinfo {pages} {4634--4639} (\bibinfo {year}
		{2016})}\BibitemShut {NoStop}%
	\bibitem [{\citenamefont {Marais}\ \emph {et~al.}(2018)\citenamefont {Marais},
		\citenamefont {Adams}, \citenamefont {Ringsmuth}, \citenamefont {Ferretti},
		\citenamefont {Gruber}, \citenamefont {Hendrikx}, \citenamefont {Schuld},
		\citenamefont {Smith}, \citenamefont {Sinayskiy}, \citenamefont {Kr{\"u}ger}
		\emph {et~al.}}]{Marais2018JRSI}%
	\BibitemOpen
	\bibfield  {author} {\bibinfo {author} {\bibfnamefont {A.}~\bibnamefont
			{Marais}}, \bibinfo {author} {\bibfnamefont {B.}~\bibnamefont {Adams}},
		\bibinfo {author} {\bibfnamefont {A.~K.}\ \bibnamefont {Ringsmuth}}, \bibinfo
		{author} {\bibfnamefont {M.}~\bibnamefont {Ferretti}}, \bibinfo {author}
		{\bibfnamefont {J.~M.}\ \bibnamefont {Gruber}}, \bibinfo {author}
		{\bibfnamefont {R.}~\bibnamefont {Hendrikx}}, \bibinfo {author}
		{\bibfnamefont {M.}~\bibnamefont {Schuld}}, \bibinfo {author} {\bibfnamefont
			{S.~L.}\ \bibnamefont {Smith}}, \bibinfo {author} {\bibfnamefont
			{I.}~\bibnamefont {Sinayskiy}}, \bibinfo {author} {\bibfnamefont {T.~P.}\
			\bibnamefont {Kr{\"u}ger}}, \emph {et~al.},\ }\bibfield  {title} {\enquote
		{\bibinfo {title} {The future of quantum biology},}\ }\href
	{https://doi.org/https://doi.org/10.1098/rsif.2018.0640} {\bibfield
		{journal} {\bibinfo  {journal} {J. R. Soc. Interface}\ }\textbf {\bibinfo
			{volume} {15}},\ \bibinfo {pages} {20180640} (\bibinfo {year}
		{2018})}\BibitemShut {NoStop}%
	\bibitem [{\citenamefont {Holland}(2014)}]{Holland2014JZ}%
	\BibitemOpen
	\bibfield  {author} {\bibinfo {author} {\bibfnamefont {R.}~\bibnamefont
			{Holland}},\ }\bibfield  {title} {\enquote {\bibinfo {title} {True navigation
				in birds: {F}rom quantum physics to global migration},}\ }\href
	{https://doi.org/https://doi.org/10.1111/jzo.12107} {\bibfield  {journal}
		{\bibinfo  {journal} {J. Zool.}\ }\textbf {\bibinfo {volume} {293}},\
		\bibinfo {pages} {1--15} (\bibinfo {year} {2014})}\BibitemShut {NoStop}%
	\bibitem [{\citenamefont {Kattnig}, \citenamefont {Solov'yov},\ and\
		\citenamefont {Hore}(2016)}]{Kattnig2016PCCP}%
	\BibitemOpen
	\bibfield  {author} {\bibinfo {author} {\bibfnamefont {D.~R.}\ \bibnamefont
			{Kattnig}}, \bibinfo {author} {\bibfnamefont {I.~A.}\ \bibnamefont
			{Solov'yov}},\ and\ \bibinfo {author} {\bibfnamefont {P.}~\bibnamefont
			{Hore}},\ }\bibfield  {title} {\enquote {\bibinfo {title} {Electron spin
				relaxation in cryptochrome-based magnetoreception},}\ }\href
	{https://doi.org/DOI https://doi.org/10.1039/C5CP06731F} {\bibfield
		{journal} {\bibinfo  {journal} {Phys. Chem. Chem. Phys.}\ }\textbf {\bibinfo
			{volume} {18}},\ \bibinfo {pages} {12443--12456} (\bibinfo {year}
		{2016})}\BibitemShut {NoStop}%
	\bibitem [{\citenamefont {Schulten}, \citenamefont {Swenberg},\ and\
		\citenamefont {Weller}(1978)}]{Schulten1978ZPC}%
	\BibitemOpen
	\bibfield  {author} {\bibinfo {author} {\bibfnamefont {K.}~\bibnamefont
			{Schulten}}, \bibinfo {author} {\bibfnamefont {C.~E.}\ \bibnamefont
			{Swenberg}},\ and\ \bibinfo {author} {\bibfnamefont {A.}~\bibnamefont
			{Weller}},\ }\bibfield  {title} {\enquote {\bibinfo {title} {A biomagnetic
				sensory mechanism based on magnetic field modulated coherent electron spin
				motion},}\ }\href {https://doi.org/10.1524/zpch.1978.111.1.001} {\bibfield
		{journal} {\bibinfo  {journal} {Z. Phys. Chem}\ }\textbf {\bibinfo {volume}
			{111}},\ \bibinfo {pages} {1--5} (\bibinfo {year} {1978})}\BibitemShut
	{NoStop}%
	\bibitem [{\citenamefont {Xie}(2022)}]{Xie2022Innovation}%
	\BibitemOpen
	\bibfield  {author} {\bibinfo {author} {\bibfnamefont {C.}~\bibnamefont
			{Xie}},\ }\bibfield  {title} {\enquote {\bibinfo {title} {Searching for unity
				in diversity of animal magnetoreception: {F}rom biology to quantum mechanics
				and back},}\ }\href
	{https://www.cell.com/the-innovation/fulltext/S2666-6758(22)00025-X}
	{\bibfield  {journal} {\bibinfo  {journal} {Innovation (Camb)}\ }\textbf
		{\bibinfo {volume} {3}} (\bibinfo {year} {2022})}\BibitemShut {NoStop}%
	\bibitem [{\citenamefont {Efimova}\ and\ \citenamefont
		{Hore}(2008)}]{Efimova2008BJ}%
	\BibitemOpen
	\bibfield  {author} {\bibinfo {author} {\bibfnamefont {O.}~\bibnamefont
			{Efimova}}\ and\ \bibinfo {author} {\bibfnamefont {P.}~\bibnamefont {Hore}},\
	}\bibfield  {title} {\enquote {\bibinfo {title} {Role of exchange and dipolar
				interactions in the radical pair model of the avian magnetic compass},}\
	}\href {https://doi.org/10.1529/biophysj.107.119362} {\bibfield  {journal}
		{\bibinfo  {journal} {Biophys. J.}\ }\textbf {\bibinfo {volume} {94}},\
		\bibinfo {pages} {1565--1574} (\bibinfo {year} {2008})}\BibitemShut {NoStop}%
	\bibitem [{\citenamefont {Maeda}\ \emph {et~al.}(2008)\citenamefont {Maeda},
		\citenamefont {Henbest}, \citenamefont {Cintolesi}, \citenamefont {Kuprov},
		\citenamefont {Rodgers}, \citenamefont {Liddell}, \citenamefont {Gust},
		\citenamefont {Timmel},\ and\ \citenamefont {Hore}}]{Maeda2008Nature}%
	\BibitemOpen
	\bibfield  {author} {\bibinfo {author} {\bibfnamefont {K.}~\bibnamefont
			{Maeda}}, \bibinfo {author} {\bibfnamefont {K.~B.}\ \bibnamefont {Henbest}},
		\bibinfo {author} {\bibfnamefont {F.}~\bibnamefont {Cintolesi}}, \bibinfo
		{author} {\bibfnamefont {I.}~\bibnamefont {Kuprov}}, \bibinfo {author}
		{\bibfnamefont {C.~T.}\ \bibnamefont {Rodgers}}, \bibinfo {author}
		{\bibfnamefont {P.~A.}\ \bibnamefont {Liddell}}, \bibinfo {author}
		{\bibfnamefont {D.}~\bibnamefont {Gust}}, \bibinfo {author} {\bibfnamefont
			{C.~R.}\ \bibnamefont {Timmel}},\ and\ \bibinfo {author} {\bibfnamefont
			{P.~J.}\ \bibnamefont {Hore}},\ }\bibfield  {title} {\enquote {\bibinfo
			{title} {Chemical compass model of avian magnetoreception},}\ }\href
	{https://doi.org/https://doi.org/10.1038/nature06834} {\bibfield  {journal}
		{\bibinfo  {journal} {Nature}\ }\textbf {\bibinfo {volume} {453}},\ \bibinfo
		{pages} {387--390} (\bibinfo {year} {2008})}\BibitemShut {NoStop}%
	\bibitem [{\citenamefont {Steiner}\ and\ \citenamefont
		{Ulrich}(1989)}]{steiner1989CR}%
	\BibitemOpen
	\bibfield  {author} {\bibinfo {author} {\bibfnamefont {U.~E.}\ \bibnamefont
			{Steiner}}\ and\ \bibinfo {author} {\bibfnamefont {T.}~\bibnamefont
			{Ulrich}},\ }\bibfield  {title} {\enquote {\bibinfo {title} {Magnetic field
				effects in chemical kinetics and related phenomena},}\ }\href
	{https://doi.org/0009-2665/89/0789-005l} {\bibfield  {journal} {\bibinfo
			{journal} {Chem. Rev.}\ }\textbf {\bibinfo {volume} {89}},\ \bibinfo {pages}
		{51--147} (\bibinfo {year} {1989})}\BibitemShut {NoStop}%
	\bibitem [{\citenamefont {Woodward}\ \emph {et~al.}(2001)\citenamefont
		{Woodward}, \citenamefont {Timmel}, \citenamefont {McLauchlan},\ and\
		\citenamefont {Hore}}]{Woodward2001PRL}%
	\BibitemOpen
	\bibfield  {author} {\bibinfo {author} {\bibfnamefont {J.}~\bibnamefont
			{Woodward}}, \bibinfo {author} {\bibfnamefont {C.}~\bibnamefont {Timmel}},
		\bibinfo {author} {\bibfnamefont {K.}~\bibnamefont {McLauchlan}},\ and\
		\bibinfo {author} {\bibfnamefont {P.}~\bibnamefont {Hore}},\ }\bibfield
	{title} {\enquote {\bibinfo {title} {Radio frequency magnetic field effects
				on electron-hole recombination},}\ }\href
	{https://doi.org/https://doi.org/10.1103/PhysRevLett.87.077602} {\bibfield
		{journal} {\bibinfo  {journal} {Phys. Rev. Lett.}\ }\textbf {\bibinfo
			{volume} {87}},\ \bibinfo {pages} {077602} (\bibinfo {year}
		{2001})}\BibitemShut {NoStop}%
	\bibitem [{\citenamefont {Cai}, \citenamefont {Caruso},\ and\ \citenamefont
		{Plenio}(2012)}]{Cai2012PRA}%
	\BibitemOpen
	\bibfield  {author} {\bibinfo {author} {\bibfnamefont {J.}~\bibnamefont
			{Cai}}, \bibinfo {author} {\bibfnamefont {F.}~\bibnamefont {Caruso}},\ and\
		\bibinfo {author} {\bibfnamefont {M.~B.}\ \bibnamefont {Plenio}},\ }\bibfield
	{title} {\enquote {\bibinfo {title} {Quantum limits for the magnetic
				sensitivity of a chemical compass},}\ }\href
	{https://doi.org/https://doi.org/10.1103/PhysRevA.85.040304} {\bibfield
		{journal} {\bibinfo  {journal} {Phys. Rev. A}\ }\textbf {\bibinfo {volume}
			{85}},\ \bibinfo {pages} {040304} (\bibinfo {year} {2012})}\BibitemShut
	{NoStop}%
	\bibitem [{\citenamefont {Stoneham}\ \emph {et~al.}(2012)\citenamefont
		{Stoneham}, \citenamefont {Gauger}, \citenamefont {Porfyrakis}, \citenamefont
		{Benjamin},\ and\ \citenamefont {Lovett}}]{Stoneham2012BJ}%
	\BibitemOpen
	\bibfield  {author} {\bibinfo {author} {\bibfnamefont {A.~M.}\ \bibnamefont
			{Stoneham}}, \bibinfo {author} {\bibfnamefont {E.~M.}\ \bibnamefont
			{Gauger}}, \bibinfo {author} {\bibfnamefont {K.}~\bibnamefont {Porfyrakis}},
		\bibinfo {author} {\bibfnamefont {S.~C.}\ \bibnamefont {Benjamin}},\ and\
		\bibinfo {author} {\bibfnamefont {B.~W.}\ \bibnamefont {Lovett}},\ }\bibfield
	{title} {\enquote {\bibinfo {title} {A new type of radical-pair-based model
				for magnetoreception},}\ }\href {https://doi.org/0006-3495/12/03/0961/8}
	{\bibfield  {journal} {\bibinfo  {journal} {Biophys. J.}\ }\textbf {\bibinfo
			{volume} {102}},\ \bibinfo {pages} {961--968} (\bibinfo {year}
		{2012})}\BibitemShut {NoStop}%
	\bibitem [{\citenamefont {Wu}\ and\ \citenamefont
		{Dickman}(2012)}]{Wu2012Science}%
	\BibitemOpen
	\bibfield  {author} {\bibinfo {author} {\bibfnamefont {L.-Q.}\ \bibnamefont
			{Wu}}\ and\ \bibinfo {author} {\bibfnamefont {J.~D.}\ \bibnamefont
			{Dickman}},\ }\bibfield  {title} {\enquote {\bibinfo {title} {Neural
				correlates of a magnetic sense},}\ }\href
	{https://doi.org/10.1126/science.1216567} {\bibfield  {journal} {\bibinfo
			{journal} {Science}\ }\textbf {\bibinfo {volume} {336}},\ \bibinfo {pages}
		{1054--1057} (\bibinfo {year} {2012})}\BibitemShut {NoStop}%
	\bibitem [{\citenamefont {Wiltschko}\ and\ \citenamefont
		{Wiltschko}(2005)}]{Wiltschko2005JCPA}%
	\BibitemOpen
	\bibfield  {author} {\bibinfo {author} {\bibfnamefont {W.}~\bibnamefont
			{Wiltschko}}\ and\ \bibinfo {author} {\bibfnamefont {R.}~\bibnamefont
			{Wiltschko}},\ }\bibfield  {title} {\enquote {\bibinfo {title} {Magnetic
				orientation and magnetoreception in birds and other animals},}\ }\href
	{https://doi.org/https://doi.org/10.1007/s00359-005-0627-7} {\bibfield
		{journal} {\bibinfo  {journal} {J. Comp. Physiol. A}\ }\textbf {\bibinfo
			{volume} {191}},\ \bibinfo {pages} {675--693} (\bibinfo {year}
		{2005})}\BibitemShut {NoStop}%
	\bibitem [{\citenamefont {Keens}, \citenamefont {Bedkihal},\ and\ \citenamefont
		{Kattnig}(2018)}]{Keens2018PRL}%
	\BibitemOpen
	\bibfield  {author} {\bibinfo {author} {\bibfnamefont {R.~H.}\ \bibnamefont
			{Keens}}, \bibinfo {author} {\bibfnamefont {S.}~\bibnamefont {Bedkihal}},\
		and\ \bibinfo {author} {\bibfnamefont {D.~R.}\ \bibnamefont {Kattnig}},\
	}\bibfield  {title} {\enquote {\bibinfo {title} {Magnetosensitivity in
				dipolarly coupled three-spin systems},}\ }\href
	{https://doi.org/https://doi.org/10.1103/PhysRevLett.121.096001} {\bibfield
		{journal} {\bibinfo  {journal} {Phys. Rev. Lett.}\ }\textbf {\bibinfo
			{volume} {121}},\ \bibinfo {pages} {096001} (\bibinfo {year}
		{2018})}\BibitemShut {NoStop}%
	\bibitem [{\citenamefont {Janitz}\ \emph {et~al.}(2022)\citenamefont {Janitz},
		\citenamefont {Herb}, \citenamefont {V{\"o}lker}, \citenamefont {Huxter},
		\citenamefont {Degen},\ and\ \citenamefont {Abendroth}}]{Janitz2022JMCC}%
	\BibitemOpen
	\bibfield  {author} {\bibinfo {author} {\bibfnamefont {E.}~\bibnamefont
			{Janitz}}, \bibinfo {author} {\bibfnamefont {K.}~\bibnamefont {Herb}},
		\bibinfo {author} {\bibfnamefont {L.~A.}\ \bibnamefont {V{\"o}lker}},
		\bibinfo {author} {\bibfnamefont {W.~S.}\ \bibnamefont {Huxter}}, \bibinfo
		{author} {\bibfnamefont {C.~L.}\ \bibnamefont {Degen}},\ and\ \bibinfo
		{author} {\bibfnamefont {J.~M.}\ \bibnamefont {Abendroth}},\ }\bibfield
	{title} {\enquote {\bibinfo {title} {Diamond surface engineering for
				molecular sensing with nitrogen—vacancy centers},}\ }\href
	{https://doi.org/10.1039/D2TC01258H} {\bibfield  {journal} {\bibinfo
			{journal} {J. Mater. Chem. C}\ }\textbf {\bibinfo {volume} {10}},\ \bibinfo
		{pages} {13533--13569} (\bibinfo {year} {2022})}\BibitemShut {NoStop}%
	\bibitem [{\citenamefont {Yang}\ \emph {et~al.}(2020)\citenamefont {Yang},
		\citenamefont {Wang}, \citenamefont {Tao}, \citenamefont {Yang},
		\citenamefont {Zhang}, \citenamefont {Ai},\ and\ \citenamefont
		{Deng}}]{Yang2020AP}%
	\BibitemOpen
	\bibfield  {author} {\bibinfo {author} {\bibfnamefont {Z.-S.}\ \bibnamefont
			{Yang}}, \bibinfo {author} {\bibfnamefont {Y.-X.}\ \bibnamefont {Wang}},
		\bibinfo {author} {\bibfnamefont {M.-J.}\ \bibnamefont {Tao}}, \bibinfo
		{author} {\bibfnamefont {W.}~\bibnamefont {Yang}}, \bibinfo {author}
		{\bibfnamefont {M.}~\bibnamefont {Zhang}}, \bibinfo {author} {\bibfnamefont
			{Q.}~\bibnamefont {Ai}},\ and\ \bibinfo {author} {\bibfnamefont {F.-G.}\
			\bibnamefont {Deng}},\ }\bibfield  {title} {\enquote {\bibinfo {title}
			{Longitudinal relaxation of a nitrogen-vacancy center in a spin bath by
				generalized cluster-correlation expansion method},}\ }\href
	{https://doi.org/https://doi.org/10.1016/j.aop.2019.168063} {\bibfield
		{journal} {\bibinfo  {journal} {Ann. Phys.}\ }\textbf {\bibinfo {volume}
			{413}},\ \bibinfo {pages} {168063} (\bibinfo {year} {2020})}\BibitemShut
	{NoStop}%
\end{thebibliography}
\end{document}